\documentclass[pra,tightenlines,10pt,notitlepage,nofootinbib,twocolumn,superscriptaddress,floatfix]{revtex4-2}

\usepackage{amsmath}
\usepackage{graphicx}
\usepackage{dsfont}
\usepackage{xcolor}
\usepackage{hyperref}
\usepackage{braket}
\usepackage[normalem]{ulem}

\hypersetup{
    colorlinks=true,       
    linkcolor=red,          
  citecolor=magenta,        
    filecolor=magenta,      
    urlcolor=cyan,           
    runcolor=cyan
}

\def\sz{\sigma^z}

\begin{document}

\title{Mitigating Errors in DC Magnetometry via Zero-Noise Extrapolation}

\author{John S. Van Dyke}
\affiliation{Johns Hopkins University Applied Physics Laboratory, Laurel, Maryland 20723, USA}

\author{Zackary White}
\affiliation{William H. Miller III Department of Physics \& Astronomy, Johns Hopkins University, Baltimore, Maryland 21218, USA}

\author{Gregory Quiroz}
\affiliation{Johns Hopkins University Applied Physics Laboratory, Laurel, Maryland 20723, USA}
\affiliation{William H. Miller III Department of Physics \& Astronomy, Johns Hopkins University, Baltimore, Maryland 21218, USA}

\date{\today}

\begin{abstract}
Zero-noise extrapolation (ZNE), a technique to estimate quantum circuit expectation values through noise scaling and extrapolation, is well-studied in the context of quantum computing. We examine the applicability of ZNE to the field of quantum sensing. Focusing on the problem of DC magnetometry using the Ramsey protocol, we show that the sensitivity (in the sense of the minimum detectable signal) does not improve upon using ZNE in the slope detection scheme. On the other hand, signals of sufficiently large magnitude can be estimated more accurately. Our results are robust across various noise models and design choices for the ZNE protocols, including both single-qubit and multi-qubit entanglement-based sensing.
\end{abstract}

\maketitle

\section{Introduction}

The field of quantum sensing has seen considerable growth in recent years, spurred by significant investment from academia, industry, and governments. Here the bane of quantum computing -- the extreme sensitivity of qubits to their external environment -- becomes an advantage, enabling precise measurements of physical quantities such as electric, magnetic, and gravitational fields, among others \cite{caves1981sensing,wineland1992sensing,holland1993sensing,bollinger1996sensing,mckenzie2002sensing,lee2002quantum,leibfried2004toward,valencia2004distant,giovannetti2004quantum,deburgh2005sensing,Schirhagl2014,Degen2017,Pirandola2018}. 

A wide variety of methods have been employed to reduce the impact of noise in quantum sensing experiments, ranging from quantum control and dynamical decoupling (DD) schemes \cite{Kotler2011,lang2015dd,sekatski2016dynamical, Ajoy2017,Frey2017,stark2017narrow,poggiali2018ctrl,tian2020ctrl,rembold2020introduction} to quantum error correction \cite{Dur2014,arrad2014qec,Herrera-Marti2015,reiter2017dissipative,zhou2018achieving,layden2018spatial,zhou2019error,rojkov2022qec}. However, existing methods face various limitations, such as the inability of DD to handle Markovian noise and the qubit and control resource requirements to implement QEC. To overcome such limitations, here we explore the application of quantum error mitigation (QEM) techniques to quantum sensing, which have been developed in the field of quantum computing. Although there has been some work along these lines \cite{Otten2019,Yamamoto2022,Kwon2023}, this direction has been largely underexplored.

The present stage of quantum computing is often referred to as the noisy intermediate-scale quantum (NISQ) era, in which quantum processors containing tens to hundreds of qubits are publicly available. The error rates achieved in such devices remain relatively high, limiting the depths of circuits and fidelity of computations that can be obtained.

In this context, QEM techniques have been developed to enhance the performance of NISQ devices in lieu of full quantum error correction (the latter requiring capabilities significantly beyond the current state-of-the-art) \cite{Cai2023}. Within the ecosystem of QEM, zero-noise extrapolation (ZNE)~\cite{Temme2017,Li2017,Giurgica-Tiron2020} is an approach that seeks to mitigate noise biasing in the estimation of expectation values by error amplification. ZNE is typically performed by executing an ensemble of circuits with scaled noise levels. Resulting estimates of a target expectation value at each noise level are then used to extrapolate to the so-called zero-noise limit to estimate the noiseless expectation value. ZNE has been successfully applied to numerous experiments on quantum hardware, including for quantum chemistry \cite{Kandala2019,GoogleAI2020} and many-body physics \cite{Chen2022,YKim2023}. Given the  success of this method, it is natural to seek other applications for it. 

We develop the cross-fertilization between ZNE and quantum sensing by taking the well-known Ramsey protocol for DC magnetometry and ZNE in an attempt to improve its performance. We examine the conditions for which ZNE yields an improvement over conventional sensing, finding that although the sensitivity is not enhanced through ZNE, the estimation accuracy can be greater for sufficiently strong fields. Our results are robust for different choices of noise models and ZNE protocols, as well as for entanglement-based sensing with GHZ states.

\begin{figure*}
    \includegraphics[width=\textwidth]{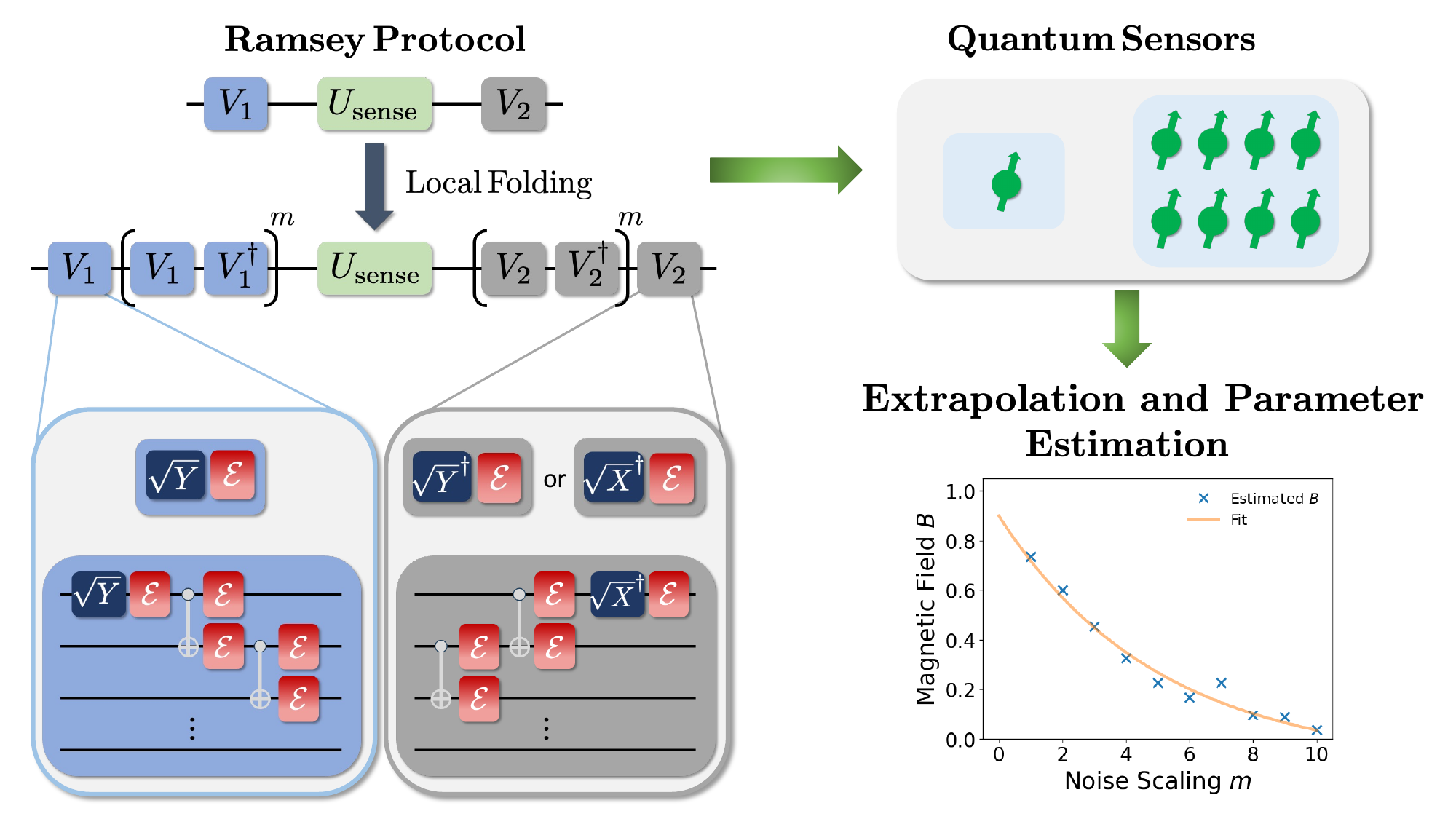}
    \caption{Overview of Ramsey-based unitary folding protocol. The generic Ramsey protocol is shown in the top-left along with a local unitary folding variant. Preparation and inversion gates $V_1$ and $V_2$, respectively, are assumed to be noisy. Their specification is dependent upon the sensing scenario. In the bottom-left, noisy preparation and inversion unitaries are shown for single-qubit and GHZ sensing. In the single qubit case, the inversion procedure depends upon the operating regime. Variance detection utilizes $V_2=\sqrt{Y}^\dagger$ gate, while slope detection corresponds to $V_2=\sqrt{X}^\dagger$. The sensor system is subject to a designated protocol, where estimates of the magnetic field $B$ are extracted as a function of the noise scaling parameter $n$. Fits to the data enable an extraction of the zero-noise limit estimate for the magnetic field strength.}
    \label{fig:znecirc}
\end{figure*}

An overview of our methodology is presented in Fig.~\ref{fig:znecirc}. ZNE noise amplification circuits are constructed from Ramsey-like protocols for both single and multi-qubit systems. System noise is assumed to be dominated by faulty control operations and thus, our protocol is based on a modified local folding technique~\cite{Schultz2022}. The sensor system is subject to a static magnetic field and driven according to the Ramsey and ZNE Ramsey-like circuits to estimate the field strength. In the case of ZNE, noise amplification is used to evaluate trends in magnetic field estimates and extract zero-noise limit predictions.

The remainder of the paper is structured as follows. Section \ref{sec:models} describes the Ramsey and ZNE protocols and noise models used in this work. Section \ref{sec:singlequbitresults} presents our numerical and analytical results for single-qubit sensing. Section \ref{sec:GHZresults} presents our results for entanglement-based sensing with GHZ states. Section \ref{sec:conclusions} summarizes our conclusions about the use of ZNE for DC magnetometry.

\section{Sensing Protocols and Noise Models \label{sec:models}}
\subsection{DC Magnetometry}
\label{subsec:dc-mag}
Quantum sensing protocols typically involve sensor initialization, a period of interaction between the signal and the sensor, followed sensor readout and estimation. In DC magnetometry, or the sensing of a static magnetic field, the canonical procedure is a Ramsey interferometry measurement. Here, we consider the Ramsey protocol applied to a single qubit sensor and its extension to an ensemble of sensors in an entangled state.

\subsubsection{Single Qubit Sensing \label{sec:singlequbitsensing}}
The Ramsey protocol is a paradigmatic method for estimating a classical parameter using a quantum system. In the typical setting, a classical external magnetic field $B$ couples to a single qubit system along its quantization axis. Realizable in a number of experimental platforms~\cite{Yoshihara2006,Taylor2008,Vengalattore2007}, the Hamiltonian effectively describing this interaction is given by
\begin{equation}
\mathcal{H} = \frac{1}{2} (B + B_0) \sz,
\end{equation}
where $B_0$ denotes a bias field from which deviations are used to infer estimates of $B$. Prior to the sensing period, the system is initialized in its ground state and subsequently prepared in a state that is maximally sensitive to the field. In the case of $\mathcal{H}$, this corresponds to preparing the equal superposition state $\ket{+}=1/\sqrt{2}(\ket{0}+\ket{1})$. State preparation is followed by a period of evolution governed by $U_{\rm sense}(t)=e^{-i \mathcal{H} t}$ where relative phase accumulation between states occurs over the sensing time $t$. The protocol is completed by reversing the state preparation and measuring in the initialization basis $\{\ket{0},\ket{1}\}$.

In the absence of a bias field, $B_0=0$, the Ramsey protocol yields a probability of measuring the $|1\rangle$ state given by
\begin{align}
    p_1 = \tfrac{1}{2}[1 + \cos (B t) ]
    \label{eqn:p1agnosticvariance}
\end{align}
so that a measurement of $p_1$ can be used to infer the magnetic field strength $B$. This regime is known as variance detection, since $p_1$ scales as $B^2$ for small $Bt$. However, the Ramsey protocol is most sensitive to weak fields in the so-called slope detection regime, where the bias field is chosen such that $B_0t=\pi/2$. This selection is associated with a reference transition probability $p_0=0.5$ from which deviations $\delta p(t)=p_1(t) - p_0$ can be related to the magnetic field strength by
\begin{equation}
\delta p(t) = \tfrac{1}{2}\sin(Bt). \label{eqn:noiselessRamsey}
\end{equation}

From a quantum circuit perspective, slope detection can be equivalently realized by setting the bias field to zero and instead altering the final unitary prior to measurement. An illustration of this protocol is shown in Fig.~\ref{fig:znecirc}, where the sensing period is bookended by unique gate operations $V_1$ and $V_2$. On the left, the state $\ket{+}$ is prepared by $V_1=\sqrt{Y}$, a $\pi/2$ rotation about the $y$-axis of the single qubit Bloch sphere. On the right, a $\sqrt{X}^\dagger$ is applied to complete the evolution. Note that this is formally equivalent to considering a bias field of $B_0=\pi/2t$ and applying $\sqrt{Y}^\dagger$. Similarly, variance detection ($B_0=0$) can be realized by maintaining the $\sqrt{Y}^\dagger$ operation. Throughout this study, we will use the Ramsey protocol as a benchmark for ZNE unitary folding procedures; see the top-left panel of Fig.~\ref{fig:znecirc}.

\subsubsection{GHZ Sensing}
Ensemble-based sensing involves using a collection of identical sensors in parallel. When the sensors are non-interacting, an ensemble of $N$ sensors acts as a collection of individual sensors. These sensors together offer a $1/\sqrt{N}$ improvement in sensitivity over a single qubit alone. This is equivalent to the classical case and is commonly known as the standard quantum limit (SQL). If instead the N qubits are placed in an entangled state, it is possible to achieve a $1/N$ (i.e., quadratic) improvement in sensitivity over the SQL. This case constitutes the well-known Heisenberg limit.

In DC magnetometry, it is common to utilize the Greenberger-Horne-Zeilinger (GHZ) state~\cite{Greenberger1990} to estimate magnetic field strengths. The protocol typically involves preparing $N$ qubits in the GHZ state $\ket{\rm GHZ}=1/\sqrt{2}(\ket{00\cdots0} + \ket{11\cdots1})$ and then allowing the system to collectively evolve according to $U(t)=e^{-i \mathcal{H}t}$ for a time $t$, where 
\begin{equation}
    \mathcal{H}=\tfrac{1}{2}(B+B_0)\sum^N_{i=1}\sz_i.
\end{equation}
Subsequently, the GHZ state preparation is reversed and the system is measured in the initialization basis. An added feature of the GHZ sensing protocol is that only one qubit needs to be measured. Note that the protocol is quite similar to the Ramsey protocol. For this reason, we will refer to it as the GHZ Ramsey protocol. A schematic for the $V_1$ and $V_2$ unitaries is shown in the bottom-left of Fig.~\ref{fig:znecirc} for slope detection; note the final unitary applied to the first qubit in $V_2$.

In the noiseless setting, the deviation in the transition probability for slope detection is given by 
\begin{equation}
\delta p(t) = \frac{1}{2}\sin(N B t).
\end{equation}
Therein lies the proportionality to $N$ which yields the well-known enhancement afforded by entanglement. Of course, the sensitivity of the GHZ state is not limited to the sensing field alone. In general, the GHZ state, like many entangled states, is strongly impacted by noise. It is this fact that typically renders entanglement-based sensing challenging in practice. Below, we will investigate ways of leveraging noise as a resource for improving estimates of static magnetic field strengths via ZNE.

\subsection{Markovian Noise Model}
In this work, errors in the sensing protocol are modeled as Markovian noise using the quantum channel formalism. We focus on errors generated during control operations. Hence, it is assumed that state preparation and inversion are faulty and sensing periods can be approximated as being noiseless. Commonly observed in atomic systems and defect centers, control-dominated errors can be a prominent noise source for quantum sensing platforms. We find that including weak noise during the sensing period does not qualitatively change our results.

Noise is modeled by the standard phase and amplitude damping channels. Each channel can be written in terms of Kraus operators (with $\alpha=P,A$ for phase and amplitude damping, respectively): 

\begin{equation}
    \mathcal{E}^{\alpha}(\rho) = E_0^\alpha \rho E^{\alpha \dagger}_0 + E_1^\alpha \rho E^{\alpha \dagger}_1,
\end{equation}
where the Kraus operators for phase damping are
\begin{align}
E_0^P = \begin{pmatrix}
1 & 0\\
0 & \sqrt{1-\lambda}
\end{pmatrix},\;
E_1^P = \begin{pmatrix}
0 & 0\\
0 & \sqrt{\lambda}
\end{pmatrix},
\label{eq:phase-damping}
\end{align}

where $\lambda$ denotes the phase damping rate, and 
\begin{align}
E_0^A = \begin{pmatrix}
1 & 0\\
0 & \sqrt{1-\gamma}
\end{pmatrix},\;
E_1^A = \begin{pmatrix}
0 & \sqrt{\gamma}\\
0 & 0
\end{pmatrix}.
\label{eq:amp-damping}
\end{align}
where $\gamma$ denotes the amplitude damping rate. Noise channels are applied locally to each qubit. Thus, in the multi-qubit case, the error channel is give by the composition $\mathcal{E}^\alpha_{1:N} = \mathcal{E}^\alpha_N \circ \cdots \circ \mathcal{E}^\alpha_1$, with $\mathcal{E}^\alpha_i$ denoting the channel applied to the $i$th qubit. Below, each error model is independently studied for Ramsey and various ZNE protocols.

\subsection{Zero-Noise Extrapolation for DC Sensing}
\label{subsec:zne}
ZNE aims to reduce noise biasing in expectation values computed on noisy quantum hardware by intentionally injecting noise into the execution of a circuit and extrapolating to the so-called zero-noise value. Noise injection can be performed in a digital manner by unitary folding, where sequences of gates are added to the circuit. Constituting identity operations in the absence of noise, these sequences enable deterministic noise scaling.

\subsubsection{Unitary Folding Protocols}
Unitary folding can be realized in a variety of ways. Local folding involves amplifying noise by folding individual gates, whereas global folding refers to folding procedures applied to the entire circuit. Here, we adapt local and global folding procedures used in quantum computing to DC magnetometry. Using the Ramsey protocol as the base noise biasing circuit, we amplify the noise due to faulty gate operations by folding state preparation, $V_1$, and inversion, $V_2$, operations to achieve local folding. We assume that the number of folds $n$ is the same for both the preparation and inversion gates. In contrast, global folding is performed by folding the Ramsey circuit, with the exception that the sensing period is not inverted. Since the system is to be freely evolving during the period of interaction with the system, it is assumed to be unaffected by the folding. An illustration of the local folding protocol is shown in Fig.~\ref{fig:znecirc}. The global folding procedure is discussed in Appendix~\ref{app:globalfold}.

\subsubsection{Fitting Procedures}
Noise scaling in ZNE is typically performed on an ensemble of circuits, each at a different noise level. Measurement outcomes of the ensemble are used to estimate expectation values of a desired observable. The resulting estimates are then fit to a functional form that ideally captures the behavior of the expectation value under the noise scaling procedure. If characteristics of the underlying noise processes are not well-understood, one must rely on fitting functions that best represent empirical trends. Commonly, this is performed via linear, Richardson, and exponential extrapolations~\cite{Giurgica-Tiron2020}. On the other hand, if one has knowledge of the noise then analytical expressions of noisy expectation value dynamics can be used to define noise-informed fitting functions. We consider both noise-agnostic and noise-informed fitting procedures.

Noise agnostic approaches rely on fitting to the estimated magnetic field strength as a function of noise level. That is, we collect $M$ estimates of the deviation in probability $\{\delta p(\lambda_1,t), \ldots, \delta p(\lambda_M,t)\}$ for different noise levels $\lambda_m = (2m+1)\lambda$. Note that we utilize the noiseless expressions for $\delta p(t)$ given in Sec.~\ref{sec:singlequbitsensing}. The magnetic field strength is then estimated for each deviation to obtain the ensemble $\mathcal{B}=\{B(\lambda_1), \ldots, B(\lambda_M)\}$. Extrapolations are ultimately performed on the ensemble $\mathcal{B}$ to estimate $B(0)$.

In contrast, noise-informed fitting leverages analytically derived expressions for the deviation in probability. It is assumed that the dominant noise source is known; however specific parameters, such as the error rates constitute unknown parameters. For Ramsey sensing, the independent variable is the sensing time. The error rate $\lambda$ and magnetic field $B$ are subsequently determined by fitting the deviation in probability $\delta p(t)$ as a function of $t$. In the case of ZNE, $B$ and $\lambda$ are determined by fitting $\delta p(\lambda_m, t)$ as a function of $m$, where the sensing time is equivalent for all $m$. Explicit expressions for $\delta p$ are dependent upon the experiment and noise model. In the subsequent section, we elaborate on the fitting functions for both Ramsey and ZNE subject to dephasing and amplitude damping.

\section{Single-qubit Sensing \label{sec:singlequbitresults}}
We now present analytical and numerical results for the performance of the ordinary Ramsey and ZNE-based protocols under Markovian phase damping and amplitude damping channels. We study the noise-agnostic and noise-informed methods described above, identifying the optimal regimes in which ZNE outperforms the simple Ramsey protocol. For our numerical results, we perform a large number of trials $n_t$ to generate statistics for the performance of each protocol on average. We note this is distinct from the number of shots $n_s$ needed to obtain a single estimate of $B$ in the Ramsey protocol.

To gauge the effectiveness of ZNE when applied to the Ramsey protocol, we introduce the metric of a {\it ZNE success probability}, $\nu$. The success probability is calculated from our simulations as the fraction of trials of the full ZNE protocol in which the extrapolated field strength is closer to the true value than estimate produced in the same run without using any folding (i.e. the ordinary Ramsey protocol, which corresponds to $n=0$).

\subsection{Phase Damping Channel}
\label{subsec:phase-damp}

\subsubsection{Noise-agnostic method}
\begin{figure}
 \includegraphics[scale=0.58]{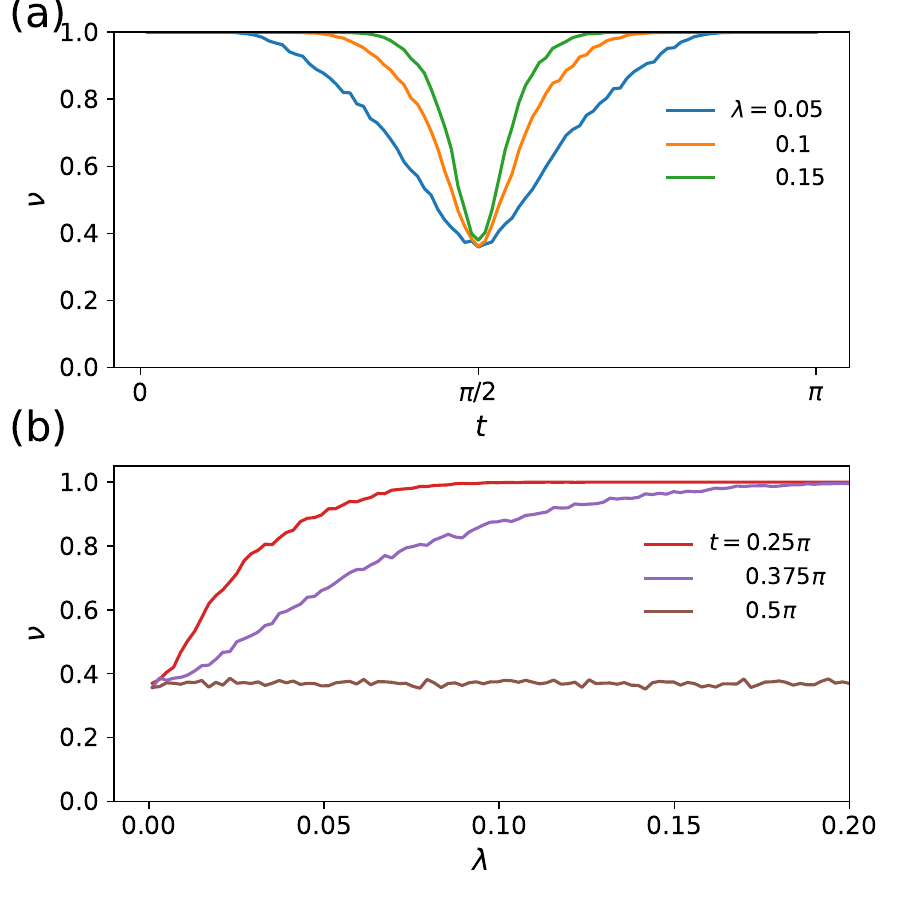}
 \caption{(a) ZNE success probability as a function of sensing time under phase damping noise, for unit $B=1$. Each individual $B$ estimate uses $n_s=10^4$ shots, and three levels of folding ($n=0,1,2$) are used for the linear extrapolation. The success probability is calculated from $n_t=5000$ trials. \label{fig:ramseyznephasedampingsensingtime}}
 \end{figure}

In the noise-agnostic method, magnetic field estimates are obtained from Eq.~\ref{eqn:p1agnosticvariance}, which does not account for the presence of noisy quantum gates. In the case of ZNE, this expression is used to infer magnetic field estimates from several circuits with different levels of folding, which are then extrapolated to obtain the ZNE estimate. We first present the ZNE success probability $\nu$ as a function of the sensing time $t$ in Fig.~\ref{fig:ramseyznephasedampingsensingtime}(a). There is a striking dip in $\nu$ centered around $t=\pi/2$, such that ZNE fails regardless of the phase damping noise strength. In contrast, increasing the noise strength leads to better ZNE performance when operating away from $t=\pi/2$ [Fig.~\ref{fig:ramseyznephasedampingsensingtime}(b)]. 

\begin{figure}
 \includegraphics[scale=0.55]{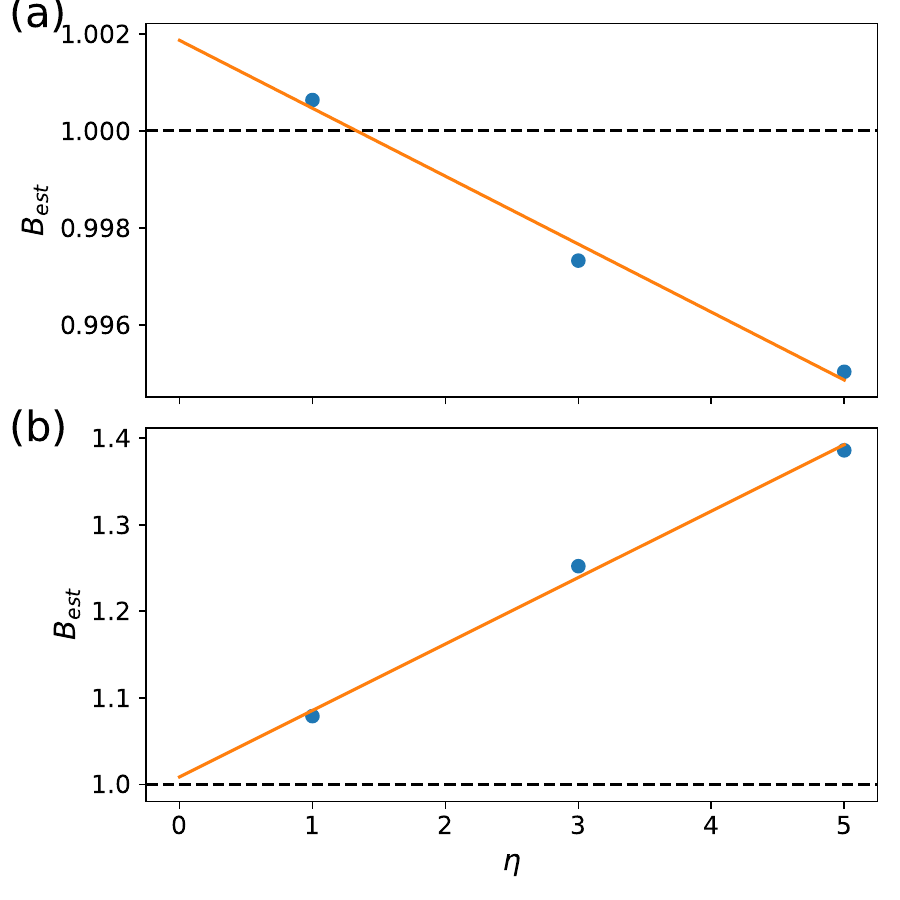}
 \caption{ZNE extrapolation of the estimated $B$ using a linear fit for (a) $t=\pi/2$; (b) $t=\pi/4$. The true value is $B=1$, the noise strength is $\lambda = 0.15$, and the individual estimates at different noise scaling values $\eta$ are obtained using $n_s=10^4$ shots each. \label{fig:ramseyznephasedampingtrials}}
 \end{figure}

The origin of this effect can be understood by looking at representative examples of the extrapolation in different regimes, where $\eta = 2m+1$ is the noise scaling factor used as the independent variable in the extrapolation. The average predicted $B$ at various levels of folding is shown in Fig.~\ref{fig:ramseyznephasedampingtrials} for two values of $t$, which are inside and outside of the low-success region in Fig.~\ref{fig:ramseyznephasedampingsensingtime}(a). At the optimal sensing time for an individual Ramsey experiment, $t=\pi/2$ [Fig.~\ref{fig:ramseyznephasedampingtrials}(a)], the estimates at different noise scaling factors do not show any systematic variation, and even occur on opposite sides of true value of $B=1$. On the other hand, when $t=\pi/4$ [Fig.~\ref{fig:ramseyznephasedampingtrials}(b)], there is a systematic overestimation of $B$, which becomes worse with increasing noise strength (similarly, $t > \pi/2$ leads to a systematic underestimation). Therefore, linear extrapolation tends to fail near the optimal $t$, while leading to improved estimates away from it.

\begin{figure}
 \includegraphics[scale=0.57]{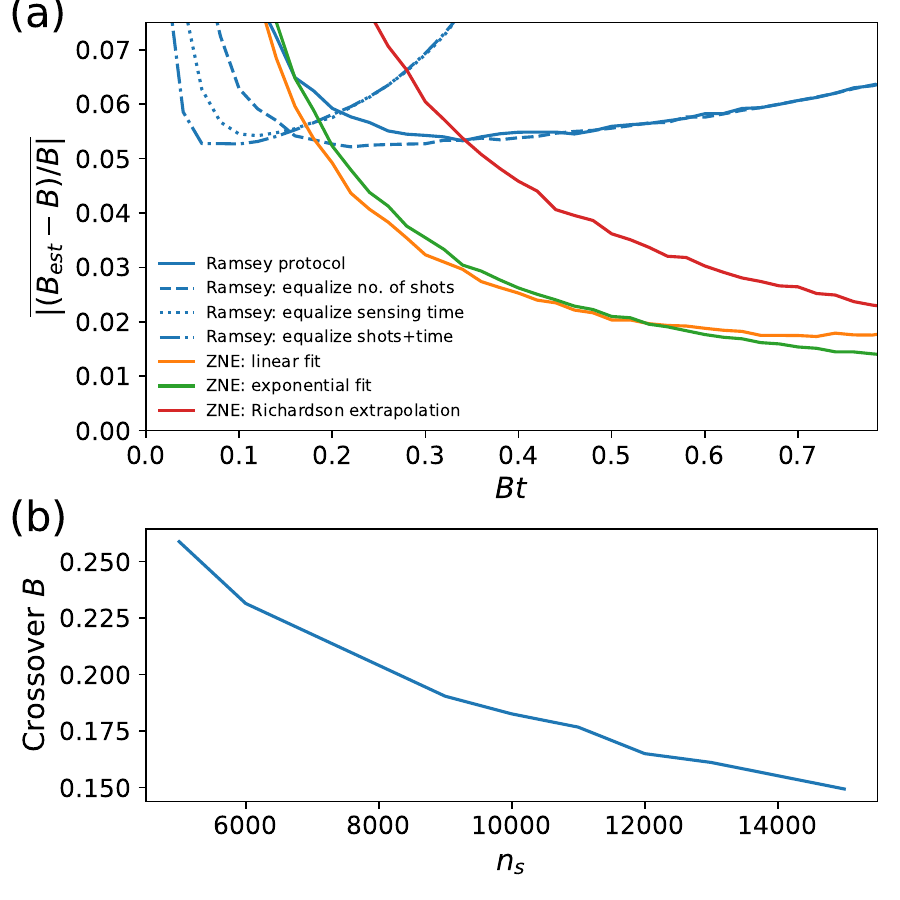}
 \caption{(a) Relative error in the estimated $B$ for the ordinary Ramsey and ZNE protocols. For the ZNE case, the individual estimates at different noise scaling factors are obtained using $n_s=10^4$ shots each. For the Ramsey protocol variants, the number of shots and/or sensing time are adjusted to equalize resources. (b) Crossover field strength from the regime in which ZNE with linear extrapolation is less accurate than the equal-shot Ramsey protocol to the regime in which it is more accurate. For both subfigures, $\lambda=0.1$, $t=1$, $n_t=5000$, and three levels of folding are used for ZNE. \label{fig:ramseyznephasedampingBerrorBcrossover}}
 \end{figure}

In practice, of course, the optimal sensing time is unknown -- one would need the value of $B$ in advance, but the latter is precisely the quantity to be determined in the experiment. As mentioned above, very often one is concerned with measuring weak fields, in which case a slope detection scheme is preferred. In Fig.~\ref{fig:ramseyznephasedampingBerrorBcrossover}(a) we consider the relative error $|(B_{est}-B)/B|$ as a function of $Bt$ for the modified slope detection Ramsey protocol, with and without ZNE. For each value of $B$, we perform a set of random trials and calculate the mean $\overline{|(B_{est}-B)/B|}$ for this set. Since the ZNE case involves running multiple Ramsey circuits, the question arises of what constitutes a fair comparison between the ordinary Ramsey and the ZNE approaches. In Fig.~\ref{fig:ramseyznephasedampingBerrorBcrossover}(a) we consider different ways of equalizing the total resources used by each method, either by setting the total number of shots or total sensing times equal (or both). Thus, if $n_f$ is the total number of circuits used by ZNE, the number of shots and/or sensing time for the Ramsey protocol is given by $n_{s,R}=n_f n_s$ and $t_{R}=n_f t$, respectively, where $n_s$ and $t$ are the values for an individual circuit in the ZNE method. Although at weak $B$ the Ramsey protocols with resource equalization outperform ZNE, there is a broad range of fields for which the relative error obtained using ZNE is lower than that of the ordinary Ramsey protocol. 

The crossover between these regimes is ultimately due to the quantum projection noise arising from the finite number of shots. Fig.~\ref{fig:ramseyznephasedampingBerrorBcrossover}(b) shows the position of the crossover as a function of the number of shots, determined for the case of the Ramsey protocol with equal total number of shots and the ZNE protocol with linear extrapolation. This indicates that the crossover moves to lower $B$ as $n_s$ is increased. By simulating the full density matrix and using it to estimate $B$ directly (equivalent to the infinite-shot limit), we find that ZNE always outperforms the standard Ramsey protocol, thus confirming the crossover is a finite-shot effect (not shown).

We note that the choice of extrapolation method for ZNE has a significant impact on the $B$ estimate error. For weak fields, Fig.~\ref{fig:ramseyznephasedampingBerrorBcrossover}(a) shows that a simple linear fit outperforms Richardson extrapolation. This can be attributed to the finite number of shots, since the latter method fits the data to a polynomial of degree one less than the number of points. As a result, random fluctuations due to shot noise are fit in this approach, leading to overfitting and poor extrapolation for weak signals. At intermediate $B$, exponential fits produce the lowest error, reflecting the increasing nonlinearity of the effects of noise with signal strength. Numerical simulations using a global folding method yield qualitatively similar results, as shown in Appendix~\ref{app:globalfold}.

The sensitivity of each protocol can be defined as the field $B$ at which the estimate error $\epsilon$ is equal to $B$. For the parameters of Fig.~\ref{fig:ramseyznephasedampingBerrorBcrossover}, we find that all variants of the Ramsey protocol have better sensitivity than the ZNE protocols. While this may seem unpromising, we note that the raw sensitivity (i.e., the minimum detectable signal) is not the only reasonable metric of performance for a quantum sensor. As seen in Fig.~\ref{fig:ramseyznephasedampingBerrorBcrossover}(a), the ZNE approach achieves a significantly better accuracy over a range of field values above the crossover point. 

The results presented above can also be understood directly from the analytic expression for the qubit excited state probability, which we derive in Appendix \ref{app:lindblad} from a Lindblad master equation approach. For $m$ levels of folding, the probability of measuring $|1 \rangle$ at the end of the Ramsey ZNE protocol is
 \begin{align}
     p_1 = \tfrac{1}{2}[1 - (1- \lambda)^{(2m+1)/2} \cos (B t)]. \label{eq:p1phasedamping}
 \end{align}

\begin{figure}
 \includegraphics[scale=0.55]{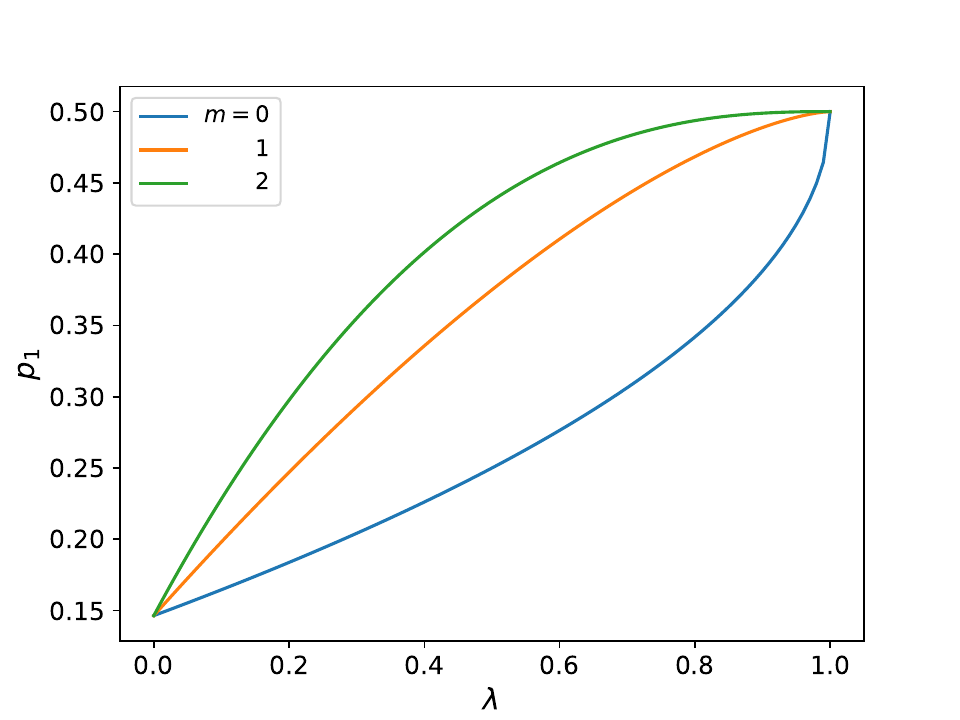}
 \caption{Excited state probability $p_1$ as a function of phase damping strength $\lambda$, for different numbers of foldings $m$, with $Bt=\pi/4$. \label{fig:ramseyznephasedampingexactp1}}
 \end{figure}

For instance, the failure of ZNE to improve upon the ordinary Ramsey protocol when $Bt = \pi/2$ (Fig.~\ref{fig:ramseyznephasedampingsensingtime}) arises from the fact that $p_1$ is independent of $m$ in this case, such that possibility of extrapolation breaks down, regardless of the form of the fitting function used. On the other hand, for generic values of $Bt$ away from $\pi/2$, the exact expression Eq.~\ref{eq:p1phasedamping} allows one to go beyond simple linear fits for ZNE extrapolation. In Fig.~\ref{fig:ramseyznephasedampingexactp1} we show the excited state probability as a function of $\lambda$ at $Bt=\pi/4$. Although a linear extrapolation in the number of foldings appears reasonable at small $\lambda$, this function clearly breaks down at intermediate and large phase damping strengths. 

\subsubsection{Noise-informed method}

While the noise-agnostic approach is expected to be useful in the absence of knowledge of the noise processes in the sensor, in the present theoretical study we are able to go further and incorporate the noise model directly into the protocols. This can be expected to further improve the performance of both the Ramsey and ZNE protocols, as the effect of noise on the transition probability can be explicitly corrected for, allowing a more accurate estimate of $B$. This is similar to the situation in quantum computing, for which characterization of noise sources enables higher fidelity operations through optimally designed protocols \cite{Schultz2022,Oda2023}. In the present case, one may use Eq.~\ref{eq:p1phasedamping} itself for the estimation of $B$, with $\lambda$ as an additional free parameter. That is, we fit the excited state probability as a function of $\eta$ to obtain values of $B$ and $\lambda$ simultaneously. This requires no explicit extrapolation to the zero noise limit, but is clearly in the spirit of ZNE, as it still uses the unitary folding method to enhance the noise. To provide a fair comparison of this two-parameter fitting approach with the ordinary Ramsey protocol, we use multiple values of $t$ for the Ramsey circuits, and fit $B$ and $\lambda$ as a function of that parameter (as is generally done in experiments that measure Ramsey fringes).

\begin{figure}
 \includegraphics[scale=0.55]{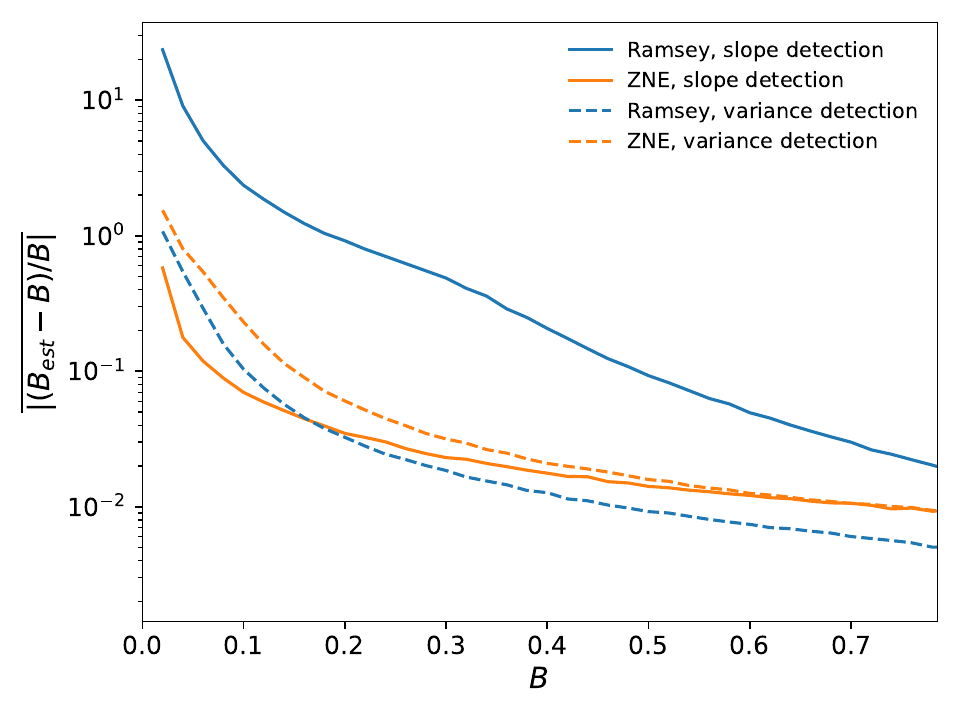}
 \caption{Relative error in the estimated $B$ for Ramsey and ZNE versions to the two-parameter fitting method. Parameters are $\lambda=0.05$, $t=1$, $n_s=20000$, $n_t=5000$. The Ramsey protocol outperforms ZNE except for weak fields. \label{fig:ramseyznephasedampingexactp1fitting}}
 \end{figure}

We compare the two approaches in a fixed total sensing time scenario. If $M$ levels of folding are employed in the ZNE approach with each circuit using a sensing time $t_Z$, the total time $Mt_Z$ is distributed among a set of Ramsey circuits whose sensing times are equally spaced, $Mt_Z = \sum_j j t_R = M(M+1)t_R/2$, where $t_R$ is the shortest sensing time in the set. For our numerical simulations, we take initial guesses for $\lambda$ and $B$ that are 99\% of their true values, to avoid trivial failures that could be removed with better fitting algorithms. Fig.~\ref{fig:ramseyznephasedampingexactp1fitting} shows the average error in $B$ as a function of the true field strength, for the Ramsey and ZNE fitting methods, using both the slope and variance detection protocols. At the lowest values of $B$, the slope detection scheme with ZNE fitting performs best, although it is quickly overtaken by Ramsey fitting using variance detection for larger fields. For low fields and the same noise strength as in Fig.~\ref{fig:ramseyznephasedampingBerrorBcrossover}, the present two-parameter methods do not provide an advantage over the noise-agnostic approach used earlier, whereas at larger fields the two-parameter Ramsey variance detection scheme has significantly lower errors than all other approaches considered. At a large phase damping strength of $\lambda=0.2$, the advantage of ZNE over Ramsey in the two-parameter fitting approach is essentially erased (not shown), suggesting that Ramsey fitting is superior in more noisy environments. 

\subsection{Amplitude Damping Channel}
In this subsection, we turn our attention to the noise that is not diagonal in the measurement basis and perform a similar analysis to Sec.~\ref{subsec:phase-damp}. We study the effects of amplitude damping using noise-agnostic and noise-informed fitting procedures. Through the former, ZNE is shown to demonstrate the ability to improve measurement outcomes in the single-qubit Ramsey protocol.

\subsubsection{Noise-agnostic method}

As in the case of phase damping, we begin our study of amplitude damping noise by considering a noise-agnostic approach, proceeding in a similar way to Sec.~\ref{subsec:phase-damp}. For a given number of foldings, multiple shots are performed, and the transition probability $p_1$ is computed. As before, a corresponding value for $B$ is computed by inverting the noiseless Ramsey expression Eq.~(\ref{eqn:p1agnosticvariance}). The resulting data are fit with either linear, exponential, or Richardson extrapolation in order to estimate the noiseless value $B_{\mathrm{est}}=B(\gamma=0)$. To compare with these ZNE estimates, we examine the case of a standard Ramsey experiment, with no additional noise mitigation applied. 

\begin{figure}
 \includegraphics[scale=0.55]{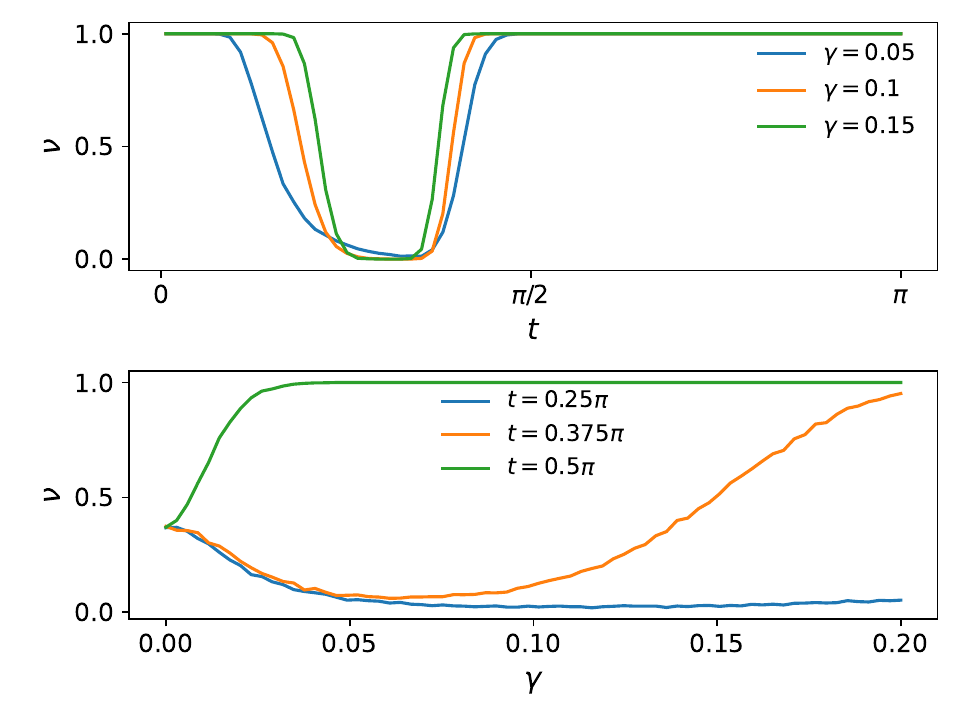}
 \caption{ZNE success probability as a function of sensing time (first panel) and noise strength (second panel) under amplitude damping noise, for unit $B=1$. Each individual $B$ estimate uses $n_s=10^4$ shots, and three levels of folding ($n=0,1,2$) are used for the linear extrapolation. The success probability is calculated from $n_t=5000$ trials.}
 \label{fig:ramseyzneamplitudedampingsensingtime}
 \end{figure}

The ZNE success probability for the amplitude damping channel is shown in Fig.~\ref{fig:ramseyzneamplitudedampingsensingtime}. Here, the true value of of the magnetic field is $B=1$. Each individual trial of $B$ uses $n_s=10^4$ shots, and three steps of folding are performed ($m=0,1,2$). As in the phase damping case a dip can be seen in which ZNE systematically fails to successfully estimate $B$. The overall success probability $\nu$ is calculated by averaging over $n_t=5000$ trials. The dip for the amplitude damping channel is not centered precisely around $t=\pi/2$, but instead localized more broadly around $t\approx 1$, with the precise location of the minimum of the dip depending weakly on the strength of the noise. As before, the dip can be understood as the result of the extrapolation procedure breaking down in this regime.  We can see this effect arising from the exact analytical expression,  
\begin{equation}
    p_1 = \frac{1}{2} \left(1-A(\gamma)-B(\gamma)\cos(B \tau_z) \right), \label{eqn:p1ADmaintext}
\end{equation}
where details on the derivation of Eq.~(\ref{eqn:p1ADmaintext}) and the functional forms of $A(\gamma)$ and $B(\gamma)$ can be found in Appendix~\ref{app:lindblad}. In Fig.~\ref{fig:ramseyzneamplitudedampingexactp1}, we plot this expression for several values of $m$ as a function of the noise strength. We use $Bt=1$ and consider $m=\{0,1,2\}$ foldings separately.

 It can be seen that higher levels of foldings produce a peak in $p_1$. This peak can be understood as a consequence of the fact that $p_1\rightarrow 0$ as $\gamma \rightarrow 1$. Increasing the number of foldings adds additional dependence on $\gamma$, which increases $p_1$ for small values of $\gamma$; however, as $\gamma$ grows, the exponential factors in $p_1$ decrease more strongly for larger $m$, leading to a peak forming. The existence of these peaks explains the breakdown of the linear extrapolations seen in the first panel of Fig.~\ref{fig:ramseyznephasedampingsensingtime}. It can be seen that in the regime corresponding to the dips in the first panel of that figure, $Bt \approx 1$ and $\gamma \approx 0.1$, the relationship between $p_1$ and $\gamma$ is highly nonlinear for $m=1$ and $m=2$. 

\begin{figure}
 \includegraphics[scale=0.55]{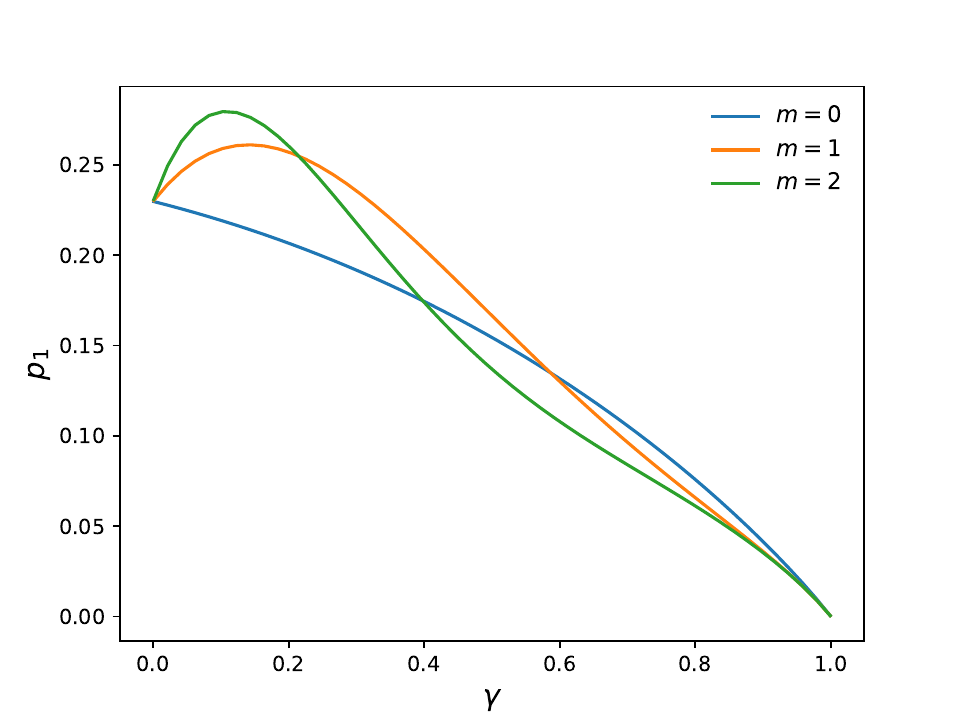}
 \caption{Exact excited state probability $p_1$ as a function of amplitude damping strength $\gamma$, for different numbers of foldings $n$, with $Bt=1$. \label{fig:ramseyzneamplitudedampingexactp1}}
 \end{figure}

\begin{figure}
 \includegraphics[scale=0.55]{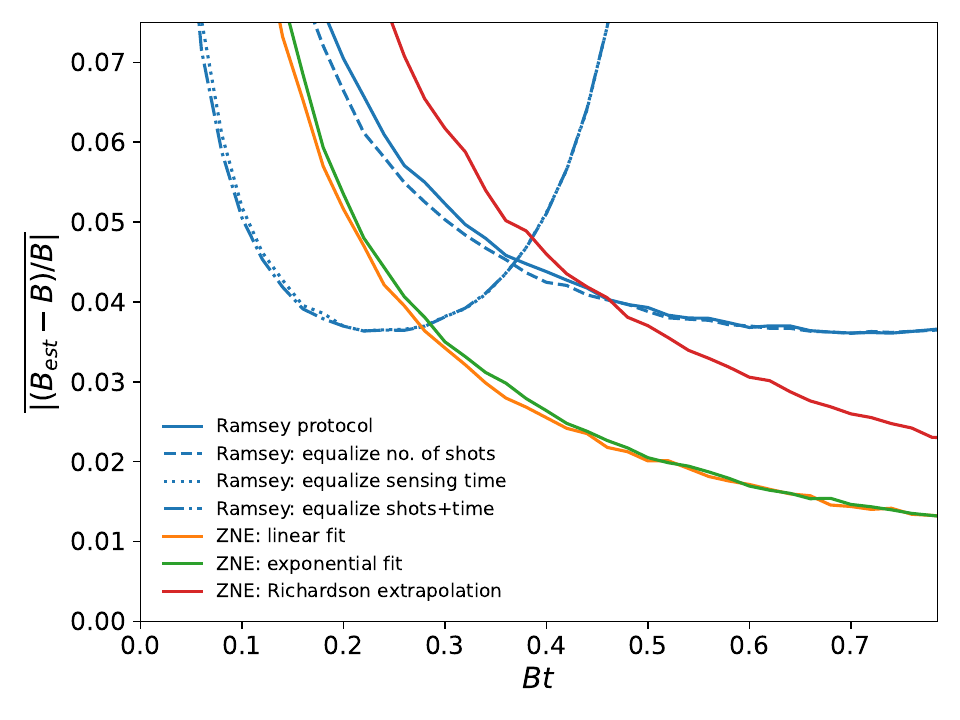}
 \caption{Relative error in the estimated $B$ for the ordinary Ramsey and ZNE protocols for noise-agnostic fits. For both cases, the individual estimates at different noise scaling factors are obtained using $n_s=10^4$ shots each. The largest number of foldings is $m=2$. For comparison, the Ramsey sequences are performed by equalizing the total number of shots, the total amount of sensing time, both, or neither. Other parameters are $\gamma=0.01$, $t=1$, $n_t=5000$.
 \label{fig:ramseyzneamplitudedampingBerroragnosticfits}}
 \end{figure}

 As before, we also study the performance of ZNE relative to an unmitigated Ramsey sequence in estimating the true value of $B$. A comparison of the relative error's for each protocol is shown in Fig.~\ref{fig:ramseyzneamplitudedampingBerroragnosticfits}. ZNE is performed using linear, exponential, or Richardson fitting on the $B$ values, with $n_s=10^4$ shots. In the amplitude damping case, $\gamma=0.01$ is needed in order to achieve relative errors of the same order of magnitude as the phase damping case. 
 
 For comparison, unmitigated Ramsey experiments are also performed, equalizing the total number of shots, the total sensing time, both, or neither (in the last case, the total number of shots for Ramsey is also $n_s=10^4$). As in the phase damping case, equalization of sensing time and/or shots is performed in order to ensure equivalent resources between Ramsey and ZNE. Once again, we take initial guesses for $\gamma$ and $B$ equal to 99\% of their true values. 
 
 For small values of $Bt$, unmitigated Ramsey sensing again outperforms any ZNE method. Also as in the phase damping case, we find that for larger values of the field, ZNE outperforms Ramsey sensing in terms of relative error in field strength. In particular, the linear fitting method gives the best results, consistently outperforming Richardson extrapolation for all values of $Bt$. This observation can once again be understood as a result of the Richardson extrapolation over-fitting to shot noise in the system. The exponential fit performs similarly to the linear fit on the axes depicted in Fig.~\ref{fig:ramseyzneamplitudedampingBerroragnosticfits}, but it is highly unstable, with relative error several orders of magnitude higher than any other approach for smaller values of $Bt$. This instability arises from the initial conditions for $\gamma$ and $B$, and demonstrates that linear extrapolation produces optimal results for ZNE while remaining more robust than alternative fitting methods. 

Similarly to the case of phase damping, we find for amplitude damping noise that Ramsey results with equalized number of shots and/or sensing time achieve greater sensitivity than any of the ZNE methods. However, we also find that for longer sensing times, all three ZNE methods are capable of achieving smaller relative errors than any of the Ramsey experiments. 

\subsubsection{Noise-informed method}

Similarly to the case of phase damping, we can consider a noise-informed approach to mitigated amplitude damping noise. Since there are other circumstances in which it is useful to learn more detailed information about the noise parameters of the system, it is worthwhile to consider whether this information could be used to increase the effectiveness of ZNE in mitigating this noise. 

In this case, we make use of Eq.~(\ref{eqn:p1ADmaintext}) to perform a two-parameter fit for both the value of the noise strength $\gamma$ and the best-estimated value of $B$. We once again consider $n_s=10^4$ shots, with $M=3$ as the maximum number of foldings. As before, we compare ZNE to a Ramsey sequence with the total sensing time equalized. We find that the relative error in both the Ramsey and ZNE approaches improve substantially, owing to the fact that more accurate estimation is being performed. However, we also note that the Ramsey sequences generally outperform ZNE for almost all values of $B$, similarly to the phase damping case. This further highlights that sufficiently detailed knowledge of the noise tends to reduce the usefulness of ZNE error mitigation in DC magnetometry. 

\begin{figure}
\includegraphics[scale=0.55]{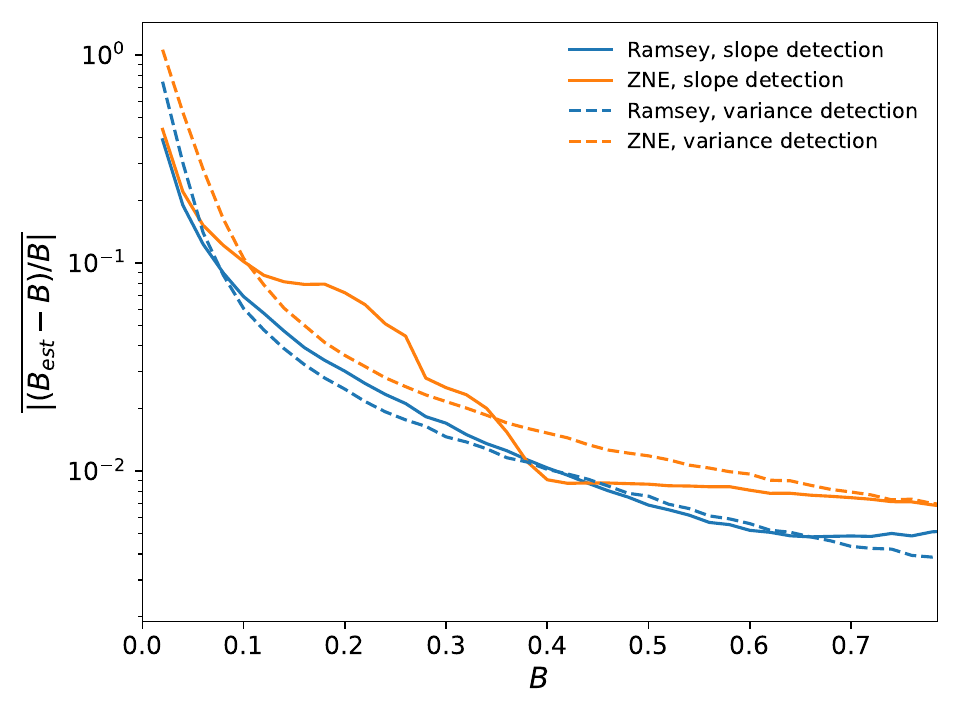}
\caption{Relative error in the estimated $B$ for the ordinary Ramsey and ZNE protocols. In this case, the extrapolation is performed as a two-parameter fit, using the exact expression for $p_1$. For both cases, the individual estimates at different noise scaling factors are obtained using $n_s=2 \times 10^4$ shots each. The largest number of foldings is $m=2$. For comparison, the Ramsey sequence is performed by equalizing the total sensing time between the two approaches. Other paramters are $\gamma=0.01$, $t=1$, $n_t=5000$.
\label{fig:ramseyzneamplitudedampingBerrorexactfit}}
\end{figure}

\section{Entanglement-Based Sensing \label{sec:GHZresults}}
Thus far, the analysis has focused on single-qubit sensors. Here, we move to the multi-qubit domain and investigate the utility of ZNE in GHZ-based DC magnetometry. The GHZ Ramsey protocol outlined in Sec.~\ref{subsec:dc-mag} is evaluated against local unitary folding in the presence of phase and amplitude damping errors resulting from faulty controls. We focus specifically on noise-agnostic fitting given that noise-informed approaches did not substantially improve ZNE in the single-qubit case.

\subsection{Phase Damping}
First, we investigate faulty gates characterized by phase damping. Each gate is followed by local phase damping error channels that are only applied to the qubits activated during the preceding gate operation. As such, the GHZ state encoding and decoding circuits are subject to noise that cascades through the circuit in a manner commensurate with the CNOT operations. Sensing periods remain noiseless as in the single qubit case.

Under these conditions, the GHZ Ramsey protocol is evaluated against ZNE using local folding as outlined in Sec.~\ref{subsec:zne}. Extrapolations are performed using $n=0,1,2$ foldings, where fits are based on the linear, Richardson, and exponential fitting functions. 

\begin{figure}[t]
    \centering
    \includegraphics[width=\columnwidth]{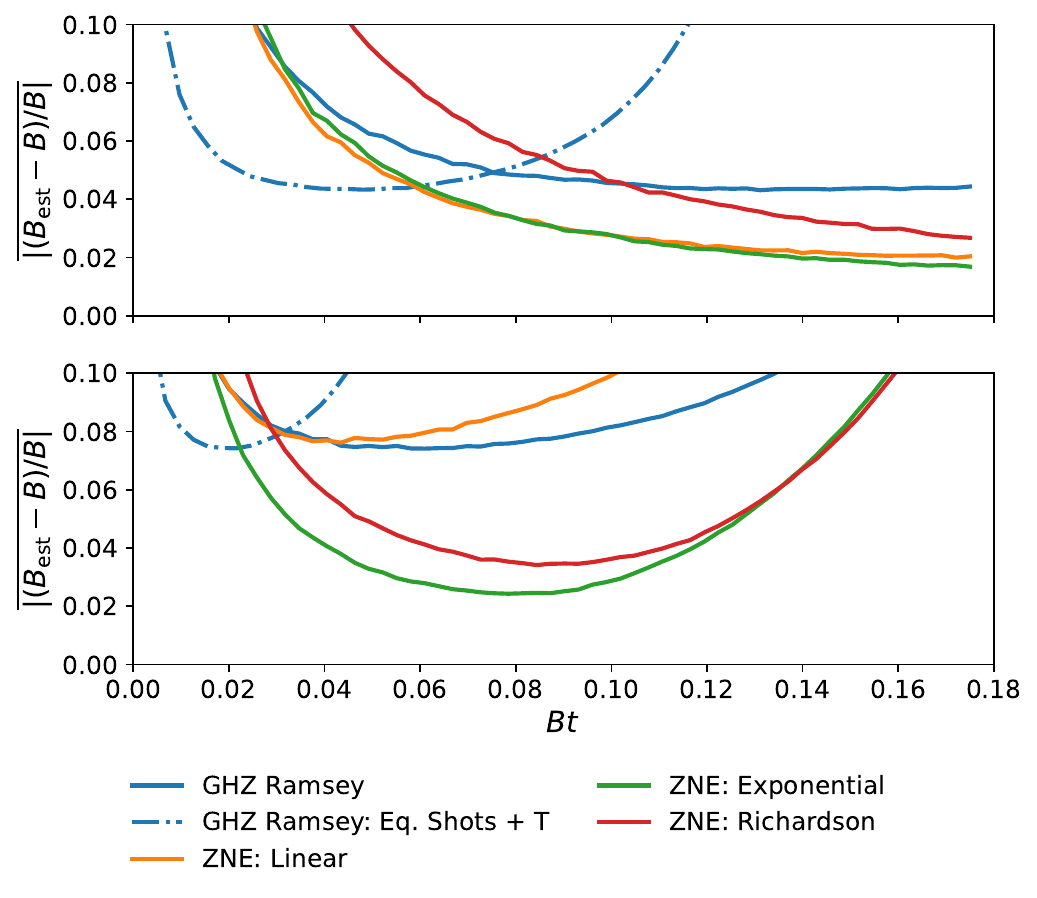}
    \caption{Relative error comparison between GHZ Ramsey and ZNE subject to phase damping. Panels (a) and (b) display results for $N=4$ and $N=8$ qubits, respectively. Numerical comparisons are performed using an error rate $\lambda=0.005$ and $10^4$ shots. ZNE extrapolations are performed using $m=0,1,2$ local foldings. Results indicate GHZ Ramsey leads to superior estimation error for small $B$, while ZNE outperforms Ramsey for larger magnetic field strengths.}
    \label{fig:ghz-phase-3}
\end{figure}

Numerical results comparing GHZ Ramsey to ZNE are displayed in Fig.~\ref{fig:ghz-phase-3}(a) and (b) for $N=4$ and $N=8$ qubits, respectively. Estimates of the average relative error are shown for $\lambda=0.005$ using $n_s=10^4$ shots and $n_t=5000$ realizations of the experiment. Note that the comparison includes two variants of the GHZ Ramsey protocol. The first denoted as GHZ Ramsey denotes the case where $t=t_{ZNE}$ and $n_{s,R}=n_s$. An additional equivalent resources comparison is made where $t=M t_{ZNE}$ and $n_{s,R}=M n_s$ as in the single qubit case. 

Both GHZ Ramsey protocols outperform ZNE for weak field strengths. Relative error rates follow a similar trend to that of the single qubit case. The equivalent resources variant is more favorable at weaker fields due to longer signal acquisition time and greater sampling resources. As in the single qubit comparison, an eventual transition is observed, where ZNE achieves lower error rates than GHZ Ramsey. This effect becomes more pronounced at lower field strengths as the number of qubits increases.

ZNE fitting procedures vary in performance depending upon the number of qubits due to the increase in noise strength. In the case of $N=4$, the noise remains relatively weak such that the linear and exponential fits are nearly equivalent. Doubling the number of qubits results in a more prominent exponential decay in the probability. As a result, the exponential fit significantly outperforms the linear case. 

Richardson extrapolation is predominately less favorable than its counterparts. For $N=4$ qubits, it consistently yields higher error rates than linear and exponential fits. Minor changes in this behavior are observed for $N=8$, where a preference towards Richardson extrapolation is found for $Bt\gtrsim0.125$. 

\subsection{Amplitude Damping}
\begin{figure}[t]
    \centering
    \includegraphics[width=\columnwidth]{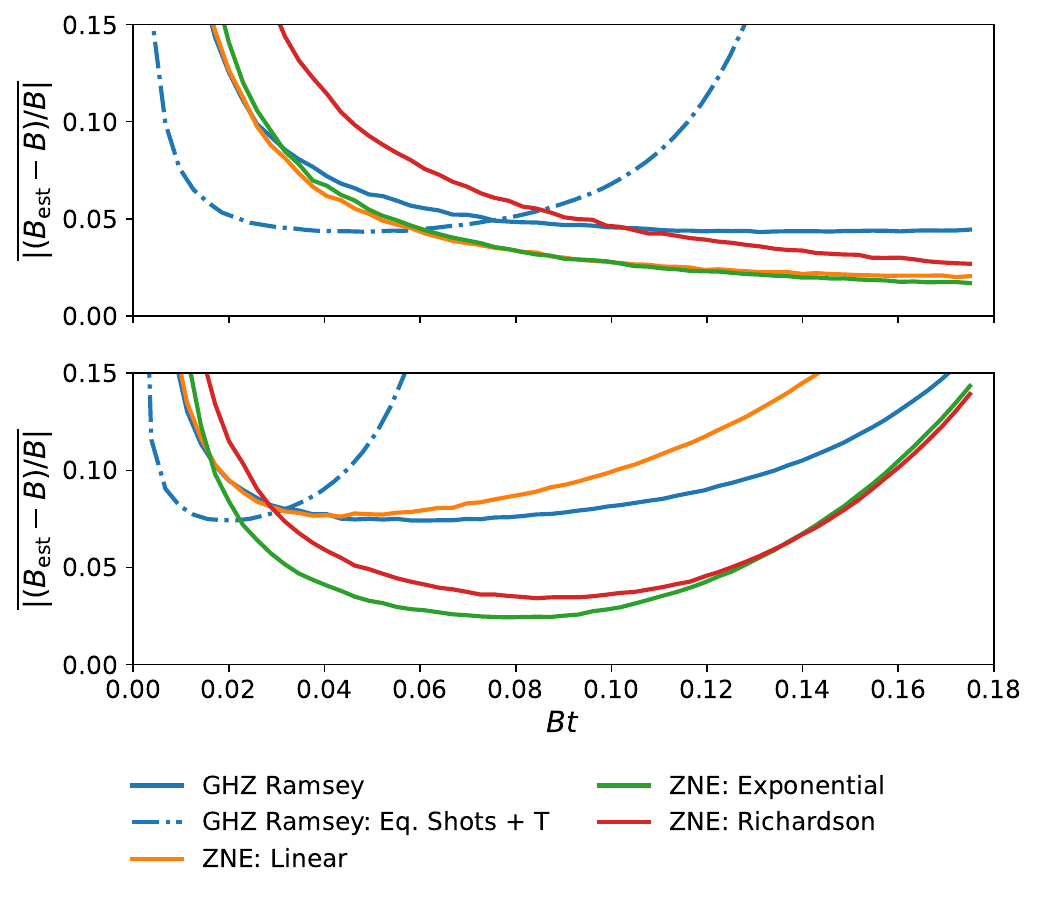}
    \caption{Relative error comparison between GHZ Ramsey and ZNE subject to amplitude damping. Numerical comparisons are performed using an error rate $\gamma=0.005$ and $10^4$ shots, with ZNE extrapolations utilizing up to $M=3$ local foldings. Results for $N=4$ and $N=8$ qubits are displayed in panels (a) and (b), respectively. Qualitatively similar to the phase damping case, GHZ Ramsey outperforms ZNE for small magnetic field strengths.}
    \label{fig:ghz-amp-3}
\end{figure}

A similar analysis is performed for faulty controls subject to amplitude damping errors. Numerical comparisons of GHZ Ramsey and ZNE utilize equivalent parameters to the phase damping case. Similarly, extrapolations are completed using up to $M=3$ foldings and for all three fitting functions. A summary of the numerical results are shown in Fig.~\ref{fig:ghz-amp-3}(a) and (b) for $N=4$ and $N=8$ qubits, respectively.

Broadly, we find that amplitude damping is more detrimental to the GHZ Ramsey-based protocols than phase damping. The latter error channel remains diagonal in the GHZ subspace and only results in loss of coherence between the states $\ket{0\cdots0}$ and $\ket{1\cdots 1}$. In contrast, amplitude damping enables evolution outside of subspace and thus, leads to further reductions in protocol performance.

Despite deviations in estimation error, the qualitative behavior for GHZ Ramsey and ZNE remain the same. Both Ramsey-like protocols outperform ZNE for small magnetic field strengths, while ZNE tends to result in improved relative error for larger field strengths. Extrapolation techniques perform similarly as well, where linear and exponential fitting yield comparable estimation error for $N=4$. Increasing the number of qubits to $N=8$ causes an enhanced exponential decay and therefore, the lowest estimation error is achieved by the exponential fit. As $B$ increases, we observe similar performance between Richardson and exponential fits.

\section{Conclusions \label{sec:conclusions}}

Rapid progress in quantum computing and quantum sensing has merited a close look at the possibility of transferring knowledge and techniques between the two fields. Here we have examined the application of ZNE to DC magnetometry, a canonical problem in quantum sensing. ZNE is an established method in NISQ quantum computing, which has demonstrated to improve the estimates of expectation values for a variety of algorithms. Applying ZNE to the well-known Ramsey protocol, we found that the nature of the noisy gate operations has a significant impact on the performance of the method. For phase damping noise and single-qubit sensing, the noise-agnostic approach to ZNE yields a worse sensitivity compared to the ordinary Ramsey protocol, when operating at the slope detection point with a finite number of shots. This ultimately arises from the failure of the extrapolation for weak fields. On the other hand, if the field is sufficiently strong, ZNE achieves a greater accuracy than the ordinary Ramsey protocol. This suggests that the method can still be useful to obtain higher-precision measurements, when the goal is not merely to sense the weakest possible signal. Similar results were also obtained under amplitude damping noise, and for entangled GHZ states. 

In the noise-informed approach to ZNE, in which $p_1$ is calculated as a function of noise strength, we find that the ordinary Ramsey protocol generally outperforms ZNE, with the exception of the weak field limit under phase damping noise, for which the performance is comparable. This further highlights that the primary virtues of ZNE are its simplicity and applicability in the absence of detailed knowledge of the noise model for the system.

Apart from the experimental demonstration of ZNE-enhanced DC magnetometry, future research directions include the analysis of ZNE applied to AC sensing protocols, and comparison with other error reduction methods for quantum sensing, such as dynamical decoupling and quantum error correction.

\begin{acknowledgments}
We thank Yasuo Oda, Paraj Titum, Leigh Norris, and Matt DiMario for helpful discussions. This work was supported by the U.S. Department of Energy, Office of Science, Office of Advanced Scientific Computing Research, Grant No. DE-SC0023196.
\end{acknowledgments}

\appendix
\section{Zero Noise Extrapolation with Global Folding}
\label{app:globalfold}
We have focused on the local folding method for ZNE in the main text, in which individual gates of the basic circuit are followed by pairs that multiply to the identity in the noiseless limit. An alternative approach known as global folding has also been used in the literature. In this case, given an initial circuit described by the unitary $U$, the circuits $(U U^\dagger)^n U$ are constructed for different values of $n$, and the expectation values from each circuit are used to extrapolate to the noiseless limit in the same manner as for local folding. 

\begin{figure}
 \includegraphics[scale=0.53]{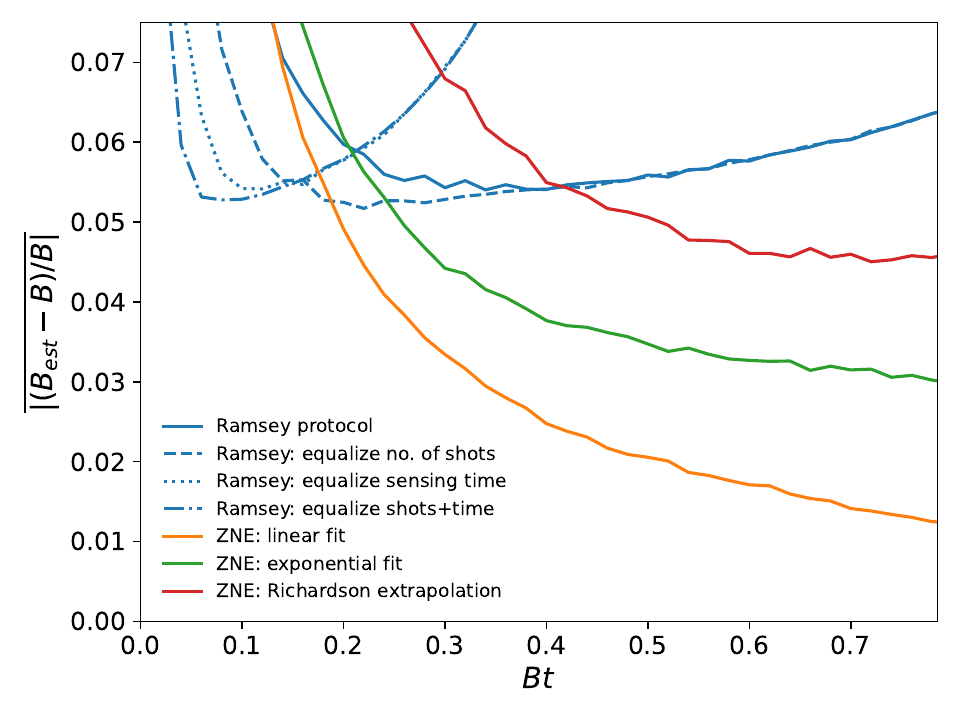}
 \caption{Relative error in the estimated $B$ for the ordinary Ramsey and ZNE protocols using global folding with the slope detection method, under phase damping noise. For the ZNE case, the individual estimates at different noise scaling factors are obtained using $n_s=10^4$ shots each. For the Ramsey protocol variants, the number of shots and/or sensing time are adjusted to equalize resources. Parameters are $\lambda=0.1$, $t=1$, $n_t=5000$, and three levels of folding are used for ZNE. \label{fig:ramseyznephasedampingglobalfold}}
 \end{figure}

Here we make a slight modification to the standard global folding method. Since the magnetic field is not under the experimenter's control, we do not change the sign of the $R_z$ gates in the Ramsey protocol when implementing the folded circuits. This implies that in the noiseless case, the total sensing time would increase with each level of folding. To prevent this, we also rescale the sensing times for the individual $R_z$ gates to be $t/(2n+1)$ in each successive ZNE circuit. Numerical simulations of the relative error for the estimate of $B$ in the presence of phase damping are shown in Fig.~\ref{fig:ramseyznephasedampingglobalfold}. The results are qualitatively similar to those obtained from local folding, with the most notable difference being the greater separation between the performance of the linear and exponential fits.

 \begin{figure}
 \includegraphics[scale=0.5]{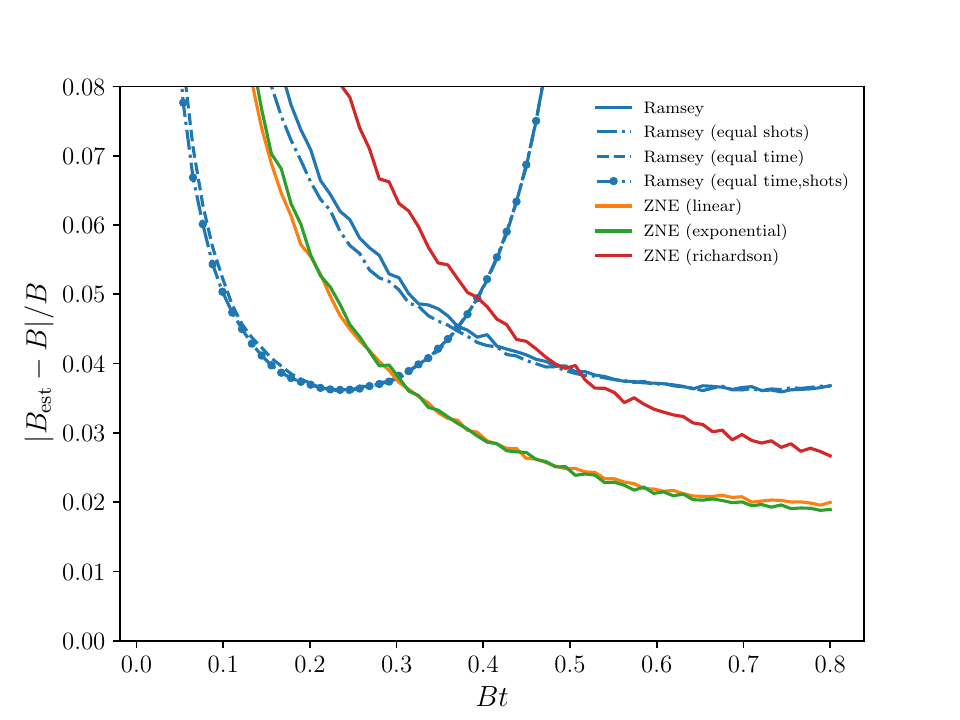}
 \caption{Relative error in the estimated $B$ for the ordinary Ramsey and ZNE protocols using global folding with amplitude damping noise and the slope detection method. For the ZNE case, the individual estimates at different noise scaling factors are obtained using $n_s=10^4$ shots each. For the Ramsey protocol variants, the number of shots and/or sensing time are adjusted to equalize resources. Parameters are $\gamma=0.01$, $t=1$, $n_t=5000$, and three levels of folding are used for ZNE. \label{fig:ramseyzneamplitudedampingglobalfold}}
 \end{figure}

\section{Analytical Results for Phase and Amplitude Damping\label{app:lindblad}}

\subsection{Local folding}

In principle, exact analytical expressions for the excitation probability $p_1$ can be computed by simply applying successive quantum channels to the system's density matrix $\rho$, corresponding to the types of noise that act during a given experiment. However, this approach is difficult to compute exactly for an arbitrary number of foldings. An alternative method is to consider a Lindblad equation approach \cite{Nielsen2000}. We start from the standard form of the Lindblad master equation 
\begin{equation}
\dot{\rho}(t) = - i [H_c, \rho] + \sum_k \gamma_k \left( L_k \rho L_k^{\dagger} - \frac{1}{2} \{ L_k^{\dagger} L_k, \rho \} \right),
\end{equation}
where the operators $L_K$ are Lindbladians and $\gamma_k$ is the decay rate for the $k$-th error channel. We denote phase damping by $\gamma_k = \Lambda$ and amplitude damping by $\gamma_k = \Gamma$. We assume constant control such that the control Hamiltonian takes the form
\begin{equation}
H_c = \frac{\theta}{2 \tau} \sigma^x 
\end{equation}
for $\theta \in \{ \pi/2, \pi \} $, and $\tau$ is the duration of the gate. Writing the control rotation rate around the $X$ axis in terms of $\omega = \theta / \tau$, (neglecting coherent under/over-rotation control noise) we have 
\begin{equation}
\begin{split}
\dot{\rho}(t) & = i \frac{\omega}{2} [ \sigma^x, \rho] + \frac{\Lambda}{2} (\sigma^z \rho \sigma^z - \rho) \\
& \; + \Gamma \left( \sigma^{+} \rho \sigma^{-} - \frac{1}{2} \{ \sigma^- \sigma^+ , \rho \} \right) .
\end{split}
\end{equation}
By decomposing the density matrix into the Bloch representation, $\rho(t) = ( I + \vec{v} \cdot \vec{\sigma} )/2 $, one obtains a set of coupled differential equations for the Bloch vector, $\vec{v}$:
\begin{equation}
\dot{\vec{v}}(t) = \mathbf{G}_c \cdot \vec{v} (t) + \vec{c} \label{eqn:vdiffyqs}.
\end{equation}
Eq.~(\ref{eqn:vdiffyqs}) is expressed in terms of the shift vector $\vec{c}$, given by
\begin{equation}
\vec{c} = ( 0, 0 , \Gamma ), 
\end{equation}
and the coupling matrix $\mathbf{G}_c$, given by
\begin{equation}
\mathbf{G}_c = \begin{pmatrix} - \left( \frac{\Gamma}{2} + \Lambda \right) & 0 & - \omega \\ 0 & - \left( \frac{\Gamma}{2} + \Lambda \right) & 0 \\  \omega & 0 & - \Gamma 
\end{pmatrix}. 
\end{equation} 
Formally, Eq.~(\ref{eqn:vdiffyqs}) has the solution 
\begin{equation}
\vec{v} (t+\tau) = e^{\mathbf{G}_c \tau} \cdot \vec{v}(t) + \left( e^{\mathbf{G}_c \tau} - 1 \right) \cdot \mathbf{G}_c^{-1} \cdot \vec{c}. \label{eqn:formalsolution}
\end{equation}
We can derive a general form for the evolution of $\vec{v}$ by iteratively applying this formal solution to the initial state of the system, alternating between gate and noise evolution operations in correspondence with the circuits in Fig.~\ref{fig:znecirc}. This ensures that the analytical calculation matches the numerical simulations, in which we apply rotation and noise gates in sequence, rather than simultaneously. In practice, this means setting either $\omega$ or $\Lambda$ ($\Gamma$) to zero when applying Eq.~(\ref{eqn:formalsolution}) to model either a dephasing (amplitude damping) noise channel or a rotation gate, respectively. 

In addition to the noise and control operators, we consider a noise-free rotation arising due to the signal that we intend to sense, $H_s = B \sigma^z$. In this case, due to the lack of noise, the coupling matrix simply takes the form 
\begin{equation}
\mathbf{G}_s = \begin{pmatrix} 0 & - B & 0 \\  B & 0 & 0 \\ 0 & 0 & 0 
\end{pmatrix}, 
\end{equation} 
with a corresponding evolution governed by 
\begin{equation}
\vec{v} (t+\tau) = e^{\mathbf{G}_s \tau_z} \cdot \vec{v}(t), \label{eqn:formalsolutionsensing}
\end{equation}
where $\tau_z$ is the total sensing time. The full solution will involve alternating between periods of control and noise governed by Eq.(\ref{eqn:formalsolution}) and periods of sensing governed by Eq.(\ref{eqn:formalsolutionsensing}).

\subsubsection{Phase damping}

For phase damping, $\vec{c}=0$, such that the action of the gates, magnetic field, and noise operations are fully captured by the application of $e^{\mathbf{G}\tau}$ on the Bloch vector. Starting from the initial state $|0\rangle = (0,\;0,\;1)^T$, we first apply the $R_x(\pi/2)$ gate, followed by the phase damping operation. The only nonzero component of the resulting Bloch vector is 
\begin{align}
    y_0 = -e^{-\Lambda t_p}
\end{align}
where the subscript ``0'' is given to denote that zero foldings have been performed thus far and $t_p$ is the time for which the continuous dephasing channel acts. A single folding corresponds to the sequence $R_x(\pi/2)$, $\mathcal{E}_{PD}$, $R_x(-\pi/2)$, $\mathcal{E}_{PD}$, which yields the nonzero Bloch vector component
\begin{align}
    y_1 = e^{-\Lambda t_p} y_0
\end{align}
This shows that successive foldings do not change the structure of the Bloch vector, but merely multiply the y-component by $e^{-\Lambda t_p}$, so that $y_{j+1} = e^{-\Lambda t_p} y_j$ or
\begin{align}
    y_{k} = -e^{-(k+1)\Lambda t_p}
\end{align}
for $k$ foldings prior to the $B$ sensing period. After the magnetic field is applied and the reverse rotation $R_x(-\pi/2)$ (with susbsequent phase damping), the Bloch vector components become
\begin{align}
    \tilde{x}_0 &= -e^{-\Lambda t_p} \sin (B t) y_k\\
    \tilde{z}_0 &= -\cos (B t) y_k
\end{align}
A similar application of foldings after the reverse rotation yields  the z-component of the final Bloch vector $\vec{v}_f$ prior to measurement,
\begin{align}
    \tilde{z}_l = e^{-l \Lambda t_p} \tilde{z}_0 = e^{-(k+l+1)\Lambda t_p} \cos (Bt)
\end{align}
The probability to measure the excited state $p_1$ is related to $\vec{v}_f$ via $1 - 2 p_1 = (\vec{v}_f)_z$, so that
\begin{align}
    p_1 = \tfrac{1}{2}[1 - e^{-(k+l+1)\Lambda t_p} \cos (Bt)]
\end{align}
This result from the Lindblad approach corresponds to a continuous dephasing channel acting over a time $t_p$. On the other hand, our numerical simulations employed discrete quantum channels using the Kraus operators
\begin{align}
E_0 = \begin{pmatrix}
1 & 0\\
0 & \sqrt{1-\lambda}
\end{pmatrix},\;
E_1 = \begin{pmatrix}
0 & 0\\
0 & \sqrt{\lambda}
\end{pmatrix} \label{eq:phasedamping}
\end{align}
We therefore connect the continuous Lindblad master equation expression to the discrete quantum channel action through the mapping

\begin{align}
    e^{-(k+l+1)\Lambda t_p} \rightarrow (1 - 2\Lambda t_p)^{\eta/2} \equiv (1 - \lambda)^{\eta/2}, \label{eqn:mapping}
\end{align}
with $\eta=2m+1$ (assuming $k=l=m$), obtaining Eq.~\ref{eq:p1phasedamping} from the main text, which agrees with the results of the numerical simulations. We note that this result can also be derived directly from the Kraus operator representation of the phase damping channel. However, for the amplitude damping case it is simpler to use the master equation approach and map to the discrete limit.

\subsubsection{Amplitude damping}
Here, we perform a similar analysis for the amplitude damping channel. To the initial state $|0\rangle$, we first apply an $R_x(\pi/2)$ rotation followed by a single instance of the amplitude damping channel $\mathcal{E}_{\mathrm{AD}}$. The resulting nonzero components of the Bloch vector are 
\begin{equation}
y_0  = - e^{-\Gamma \tau/2}, \qquad z_0  = 1 - e^{-\Gamma \tau}. \label{eqn:y0z0exact} 
\end{equation}
The subscript $0$ corresponds to the fact that $0$ foldings have been performed at this stage. Next, we apply an $R_x(\pi/2)$ rotation, followed by $\mathcal{E}_{\mathrm{AD}}$, followed by $R_x(-\pi/2)$, followed by another $\mathcal{E}_{\mathrm{AD}}$. This procedure corresponds to a single folding, and leaves the nonzero components of the Bloch vector as 
\begin{align}
y_1 & =  e^{-3\Gamma \tau /2} y_0 + e^{-\Gamma \tau /2} - e^{-3\gamma \tau/2}  \\
z_1 & =  e^{-3\gamma \tau /2} z_0 + 1 - e^{-\Gamma \tau} . \label{eqn:y1z1fullgamma}
\end{align}
Here, the subscript $1$ denotes the single folding that has been performed. Equation~(\ref{eqn:y1z1fullgamma}) shows that the effect of each folding on the Bloch vector is periodic, and thus, a recursive relationship exists between $y_j$ and $y_{j+1}$ for any integer $j$ (and similarly for $z_j$ and $z_{j+1}$). As a result, we can write 
\begin{align}
y_j & = e^{- 3j \Gamma \tau /2 } y_0 + y_0 \left[ e^{- \Gamma \tau}-1 \right] \Theta (j-1) \sum_{k=0}^{j-1} e^{-3k \Gamma \tau/2} \label{eqn:justyj} \\ 
z_j & = z_0 \sum_{k=0}^j e^{-3k\Gamma \tau/2}.   \label{eqn:yjzj}
\end{align}
Here, $\Theta(x)$ is the Heaviside function, with $\Theta(x)=0$ for $x<0$ and $\Theta(x)=1$ for $x \geq 0$. Its presence reflects the fact that the second term in Eq.~(\ref{eqn:justyj}) only appears for a folded circuit. 

The next step in the evolution is the sensing period itself, $R_z (BT)$. After this period, the components of the Bloch vector are
\begin{equation}
\begin{split}
\tilde{x}_0&  = - y_n \sin(B \tau_z) \\
\tilde{y}_0 & =  z_n e^{-\Gamma \tau/2}\\
\tilde{z}_0 & = 1 - e^{-\Gamma \tau} - y_n \cos (B \tau_z) e^{-\Gamma \tau},
\end{split}
\end{equation}
where $y_n$ and $z_n$ are drawn from Eq.~(\ref{eqn:yjzj}) with $j=n$. To complete the evolution, we perform a rotation $R_x(-\pi/2)$ followed by a noise channel $\mathcal{E}_{\mathrm{AD}}$. Afterwards, we can fold the resulting Bloch vector, performing sequences of operations of the form $R_x(-\pi/2)$, $\mathcal{E}_{\mathrm{AD}}$, $R_x(\pi/2)$, and $\mathcal{E}_{\mathrm{AD}}$. Identifying a similar recursion relation, we can compute the final component of the Bloch vector $v_z$, after an arbitrary number of foldings: 
\begin{equation}
\begin{split}
v_z^{(n,m)} & = (1-e^{-\Gamma \tau}) \times \\
& \; \bigg[e^{-3 m \Gamma \tau /2} \left(1 - g_n \cos B \tau_z \right) + g_m \bigg] \\ 
& \; + g_n e^{-3(1+n+m) \Gamma \tau/2}  \cos B \tau_z  . \label{eqn:vznm}
\end{split}
\end{equation}
For compactness, we have defined
\begin{equation}
g_j =  \frac{ 1 - e^{-3 j \Gamma \tau/2}}{1 - e^{-3\Gamma \tau/2}} \label{eqn:geometricsum}
\end{equation}
for $j \in \{n,m\}$, where $n$ and $m$ correspond to the number of foldings in the first control period (before the sensing period) and the second (after the sensing period), respectively. Throughout the paper, we restrict our attention to the case of $m=n$. Allowing the values to differ does not qualitatively change the performance of ZNE relative to standard Ramsey interferometry. Note also that the Heaviside function $\Theta(j-1)$ disappears from Eq.~(\ref{eqn:vznm}), because the formal solution of the geometric sum Eq.~(\ref{eqn:geometricsum}) vanishes for $m=0$, even if the sum itself is only physical meaningful for $m \geq 1$. From Eq.~(\ref{eqn:vznm}) we can find $p_1$ as 
\begin{equation}
    p_1 = \frac{1}{2} \left(1-A(\Gamma)-B(\Gamma)\cos(B \tau_z) \right), \label{eqn:p1AD}
\end{equation}
where
\begin{align}
    A(\Gamma) & = \left(1-e^{-\Gamma \tau} \right) \left(e^{-3 m \Gamma \tau /2}  + g_m \right)  \\
    B(\Gamma) & =  e^{-3(m+1Noisepor)\Gamma \tau/2} \left(  e^{-3m \Gamma \tau/2} + \left[ e^{- \Gamma \tau } - 1\right] g_m \right) .
\end{align}

We implement the amplitude damping operations via an amplitude damping channel, which is given in terms of a decay probability $\gamma$. However, in the expression above, $\Gamma$ corresponds to a decay probability \textit{rate}, with units of inverse time. To relate these quantities, we use 
\begin{equation}
    \gamma = 1 - e^{- \Gamma \tau} \label{eqn:mappingAD}
\end{equation}
This relationship can be derived by considering the application of a series of amplitude damping channels, each of which acts for time $\delta t$, and taking the limit $\delta t \rightarrow 0$ \cite{Preskill2015}.

\subsection{Global folding}

The method discussed in the previous section gives exact results in the case of local folding. However, for global folding, the computations become more involved due to the repetitions of the sensing period during successive foldings. Nevertheless, global folding exhibits periodicity that is conducive to well-established techniques commonly employed in the analysis of periodic pulse sequences~\cite{Levitt1992,allard1998complete,helgstrand2000simulations}; namely, average Liouvillian theory (ALT)~\cite{Ghose2000,cheng2004alt}. Here, we utilize ALT to obtain approximate analytical expressions for global folding under both phase and amplitude damping.

ALT is particularly powerful when a sequence of control pulses is applied periodically. A generic solution to the Lindblad equation under this assumption can be written
\begin{equation}
\rho(t) = \exp(- \mathcal{L} t_n ) \prod_{j=1}^{n-1} [ R_j \exp( - \mathcal{L} t_j ) ] \rho(0)  \label{eqn:rhotgen}
\end{equation} 
where $R_j$ is a superoperator corresponding to the $j$th pulse in the sequence and $t_j$ is the time interval associated with the $j$th pulse, such that 
\begin{equation}
    t = \sum_{i=1}^n t_j.
\end{equation} 
It can be shown that, by defining modified Liouvillian superoperators for each time interval
\begin{equation}
    \mathcal{L}'_j = R_{n-1} R_{n-2} \cdots R_j \mathcal{L} R_j^{-1} \cdots R_{n-2}^{-1} R_{n-1}^{-1}, \label{eqn:liouvilleprime}
\end{equation} 
one can express the dynamics of a periodic evolution in terms of an \textit{average} evolution
\begin{equation}
\rho(t) = \exp (\mathcal{L}_{\mathrm{av}} t ) \rho'(0) 
\end{equation} 
with
\begin{equation}
\mathcal{L}_{\mathrm{av}} = \sum_{j=1}^{\infty} \mathcal{L}^{(j)}
\end{equation} 
The first two terms
\begin{equation}
\begin{split}
\mathcal{L}^{(1)} & = \frac{1}{t} \sum_j \mathcal{L}'_j t_j  \\ 
\mathcal{L}^{(2)} & = - \frac{1}{2t} \sum_{j>k} [ \mathcal{L}'_j t_j, \mathcal{L}'_k t_k]  
\label{eqn:averageLiouvillianterms}
\end{split}
\end{equation} 
are derived from the Magnus expansion~\cite{Magnus1954}. We apply this formalism to the evolution of a single Ramsey sequence with noisy $\tilde{R}_y(\pm \pi/2)$ gates:
\begin{equation}
    U = \tilde{R}_y\left(-\frac{\pi}{2}\right) e^{-i \mathcal{L}_H t} \tilde{R}_y \left(\frac{\pi}{2}\right), \label{eqn:URamseynoisy}
\end{equation}
where $\mathcal{L}_H$ is the Liouvillian for the sensing period. Furthermore, we model the noisy gates in terms of noiseless $R_y(\pm \pi/2)$ gates followed by the action of the noise (in line with the approach used during the simulation) via:
\begin{equation}
    \tilde{R} \left(\pm \frac{\pi}{2} \right)  = e^{-\mathcal{L} \delta T} R_y\left(\pm \frac{\pi}{2} \right). \label{eqn:RSuzukiTrotter}
\end{equation} 
Here, $\delta t$ is the gate time, and $\mathcal{L}$ is the Liouvillian associated with the action of the noise. We can expand Eq.~(\ref{eqn:URamseynoisy}) using Eq.~(\ref{eqn:RSuzukiTrotter}). By inserting pairs of $R_{\pm}^{-1} R_{\pm}$ superoperators, and recalling that $R_{\pm}^{-1}=R_{\mp}$, we find 
\begin{equation}
    U = e^{- \mathcal{L}'_2 \delta T } e^{-i \mathcal{L}'_H T} e^{- \mathcal{L}'_1 \delta T} \label{eqn:URamseyLs} 
\end{equation}
where 
\begin{equation}
\begin{split}
    \mathcal{L}'_1 & = R_y^{-1} \mathcal{L} R_y \\
    \mathcal{L}'_2 & = \mathcal{L} \\ 
    \mathcal{L}'_H & = R_1^{-1} \mathcal{L}_H R_1 . \label{eqn:liouvilleprimes}
\end{split}
\end{equation} 

We can immediately expand Eq.~(\ref{eqn:URamseyLs}) according to Eq.~(\ref{eqn:averageLiouvillianterms}) for any given type of noise. 

\subsubsection{Phase damping}

In the case of phase damping, the LME is given by
\begin{equation}
\frac{d}{dt} \rho_S = - i [ H_c , \rho_S ] + \frac{\Lambda}{2} (\sigma^z \rho \sigma^z - \rho)
\end{equation} 
where $H_c = (\omega/2) \sigma^y$. Vectorizing this evolution gives rise to the corresponding Liouvillians
\begin{equation}
\begin{split}
\mathcal{L} & = \begin{pmatrix}  0 & 0 & 0 & 0 \\ 0 & -\Lambda & 0 & 0 \\ 0 & 0 & -\Lambda  & 0 \\ 0 & 0 & 0 & 0
\end{pmatrix} \\
\mathcal{L}_H & = \begin{pmatrix} 0 & 0 & 0 & 0 \\ 0 & - i B & 0 & 0 \\ 0 & 0 & i B & 0  \\ 0 & 0  & 0& 0 \end{pmatrix} . 
\end{split}
\end{equation} 
Acting with appropriately vectorized $R_y(\pm \pi/2)$ operators gives 
\begin{equation}
\begin{split}
\mathcal{L}_1' & = \begin{pmatrix} - \frac{\Lambda}{2} & 0 & 0 & \frac{\Lambda}{2} \\ 0 & - \frac{\Lambda}{2} & \frac{\Lambda}{2} & 0 \\ 0 & \frac{\Lambda}{2}  & -\frac{\Lambda}{2} & 0 \\ \frac{\Lambda}{2} & 0 & 0 & - \frac{\Lambda}{2}\end{pmatrix} \\
\mathcal{L}_2' & = \begin{pmatrix} 0 & 0 & 0 & 0 \\ 0 & -\Lambda & 0 & 0 \\ 0 & 0 & -\Lambda  & 0\\ 0 & 0& 0& 0 \end{pmatrix} \\
\mathcal{L}'_H & = \begin{pmatrix} 0 & - \frac{i B}{2} & \frac{i B}{2}  & 0 \\ - \frac{i B}{2}  & 0 & 0 & \frac{i B}{2}  \\ \frac{i B}{2}  & 0 & 0 & - \frac{i B}{2}  \\ 0 & \frac{i B}{2}  & -\frac{i B}{2}  & 0 \end{pmatrix} . \label{eqn:LprimesPD}
\end{split}
\end{equation} 
Next, we can expand Eq.~(\ref{eqn:averageLiouvillianterms}) using Eq.~(\ref{eqn:LprimesPD}). To first order, the result is 
\begin{equation}
\mathcal{L}^{(1)} t = \frac{1}{2} \begin{pmatrix} - \Lambda \delta t & - i BT & i BT & \Lambda \delta t \\ -i BT & -3 \Lambda \delta t & \Lambda \delta t & i BT \\ i BT & \Lambda \delta t & -3 \Lambda \delta t & - i BT \\ \Lambda \delta t & i BT & - i BT & - \Lambda \delta t \end{pmatrix} , \label{eqn:PDoneaveragesequence}
\end{equation}
where $t=T+2\delta t$ is the total time of a single pulse sequence. Each folding accrues an additional factor of Eq.~(\ref{eqn:PDoneaveragesequence}), therefore for $m$ foldings the corresponding transition probability can be computed by acting on the initial state $\ket{0}$ with $ \exp[ (m+1)\mathcal{L}^{(1)} t]$. We note here that, in order to relate the dephasing rate obtained by doing so with the dephasing probability obtained from the quantum channel approach, we use 
\begin{align}
    e^{-\Lambda t_p} \rightarrow (1 - \lambda)^{1/2},
\end{align}
which is derived from Eq.~(\ref{eqn:mapping}) by setting $n=1$. This approach is necessary because ALT considers only a single control sequence, with the effect of multiple repetitions accounted for entirely by the additional overall factor in of $m+1$.

\begin{figure}
 \includegraphics[scale=0.5]{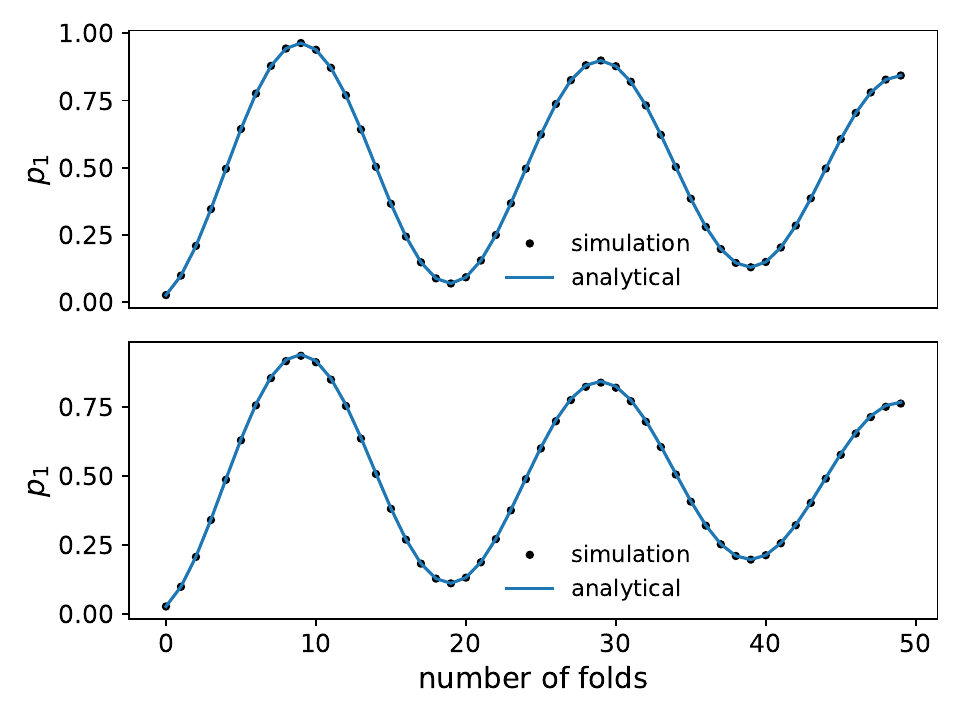}
 \caption{Transition probability $p_1$ vs. number of foldings $k$ for global folding. The top panel depicts the results for the phase damping channel, while the bottom depicts results for the amplitude damping channel. The dots depict results of the numerical simulation, while the solid line shows the analytical results. In the case of phase damping, a first-order average Liouvillian calculation is sufficient for thet noise strength of $\Lambda = 0.05$. In the case of amplitude damping, the second-order average Liouvillian contributions must be considered for a noise strength of $\Gamma=0.01$. In both panels, $B=1$ and $T=\pi/10$ have been chosen for illustrative purposes. Because the global folding procedure causes the sensing periods to repeat, the total sensing time at each level of folding $m$ is equal to $mT$. \label{fig:globalfoldinganalytic}}
 \end{figure}

In Fig.~\ref{fig:globalfoldinganalytic}(a) we depict the transition probability under global folding as a function of number of foldings $k$ for the numerical simulation and the first-order average Liouvillian theory calculation, demonstrating excellent agreement for phase damping noise.

\subsubsection{Amplitude damping}

We approach amplitude damping noise in a similar fashion as phase damping noise. The LME takes the form 
\begin{equation}
\dot{\rho}(t) = - i \frac{\omega}{2} [ H_c, \rho_S] + \gamma \left( \sigma^{+} \rho \sigma^{-} - \frac{1}{2} \{ \sigma^- \sigma^+ , \rho \} \right) .
\end{equation}
Clearly, the non-dissipative term is identical to the phase damping case. For the dissipative terms, we have 
\begin{equation}
    \mathcal{L} = \begin{pmatrix} 0 & 0 & 0 & \gamma \\ 0 & - \frac{\gamma}{2} & 0 & 0 \\ 0 & 0 & - \frac{\gamma}{2} & 0 \\ 0 & 0 & 0 & - \gamma \end{pmatrix}
\end{equation}
and
\begin{equation}
\begin{split}
\mathcal{L}_1' & = \begin{pmatrix} -\frac{\gamma}{4} & 0 & 0 & \frac{\gamma}{4} \\ - \frac{\gamma}{2} & - \frac{3\gamma}{4} & -\frac{\gamma}{4} & - \frac{\gamma}{2} \\ - \frac{\gamma}{2} & - \frac{\gamma}{4} & - \frac{3\gamma}{4} & - \frac{\gamma}{2} \\ \frac{\gamma}{4} & 0 & 0 & - \frac{\gamma}{4} \end{pmatrix} \\
\mathcal{L}_2' & = \begin{pmatrix} 0 & 0 & 0 & \gamma \\ 0 & -\frac{\gamma}{2} & 0 & 0 \\ 0 & 0 & -\frac{\gamma}{2}  & 0\\ 0 & 0& 0 & -\gamma \end{pmatrix}. \label{eqn:LprimesAD}
\end{split}
\end{equation}
From Eq.~(\ref{eqn:LprimesAD}) we derive 
\begin{equation}
    \mathcal{L}^{(1)} t = \frac{1}{2} \begin{pmatrix} - \frac{\gamma \delta t}{2} & - i BT & i BT & \frac{5 \gamma \delta t}{2} \\ -i B T - \gamma \delta t & -\frac{5 \gamma \delta t}{2} & - \frac{\gamma \delta t}{2} & i B T - \gamma \delta t \\ i B T - \gamma \delta t & - \frac{\gamma \delta t}{2} & - \frac{5 \gamma \delta t}{2} & - i B T - \gamma \delta t \\ \frac{\gamma \delta t}{2} & i B T & - i B T & - \frac{5 \gamma \delta t}{2} \end{pmatrix}.
\end{equation}

For larger values of $\Gamma$, amplitude damping requires a second-order contribution, which can be computed as 
\begin{equation}
\mathcal{L}^{(2)} t  = \frac{1}{8} \begin{pmatrix} \gamma^2 \delta t^2 & i B T \gamma \delta t & -i B T \gamma \delta t & \gamma^2 \delta t^2 \\ \gamma \delta t f_+ & 0 & 0 & \gamma \delta t g_+ \\ \gamma \delta t f_- & 0 & 0 & \gamma \delta t g_- \\ - \gamma^2 \delta t^2 & - i B T \gamma \delta t & i B T \gamma \delta t & - \gamma^2 \delta t^2 \end{pmatrix},
\end{equation}
where we have defined 
\begin{equation}
\begin{split}
    f_{\pm} & \equiv \gamma \delta t \pm i B T  \\
    g_{\pm} & \equiv \gamma \delta t \pm 3 i B T 
\end{split}
\end{equation}
for compactness. Once again, the transition probabilities are computed by acting on the state $\ket{0}$ with the operation $\exp[(m+1)(\mathcal{L}^{(1)}t+\mathcal{L}^{(2)}t)]$. Similarly to the case of phase damping, we relate the decay probability $\gamma$ to the decay rate $\Gamma$ using Eq.~(\ref{eqn:mappingAD}). Fig.~\ref{fig:globalfoldinganalytic}(b) illustrates the agreement between the analytic and numerical results for the amplitude damping case.

\bibliography{bibramseyzneDC}

\end{document}